\documentclass[conference]{IEEEtran}

\usepackage[usenames,dvipsnames,table]{xcolor}

\usepackage{graphicx} 
\usepackage{booktabs} 
\usepackage{svg} 

\usepackage{multirow} 
\usepackage{varwidth} 

\usepackage{amsthm}
\theoremstyle{definition}

\usepackage{graphicx}
\usepackage[
  colorlinks=purple,
  linkcolor=purple,
  citecolor=purple,
  urlcolor=purple,
  pdfauthor={},
  pdftitle={},
  pdfsubject={},
  pdfkeywords={},
  bookmarks=false,
]{hyperref}
\hypersetup{colorlinks=true}

\newif\ifnotes
\notestrue 

\usepackage[
  sorting=none,
  style=numeric-comp,
  backend=biber,
  isbn=false,
  url=true,
  doi=false,
  url=false,
  mincrossrefs=100,
  maxnames=12,
  firstinits=true
]{biblatex}

\AtEveryBibitem{\clearname{editor}}

\title{SoK: Microservice Architectures from a Dependability Perspective}

\author{
  \IEEEauthorblockN{
  	Dāvis Kažemaks, Jérémie Decouchant
  }
  \IEEEauthorblockA{
    Delft University of Technology \\
    d.kazemaks@student.tudelft.nl, j.decouchant@tudelft.nl
    }
}

\date{}

\DeclareUnicodeCharacter{0307}{XXXXXXXXXXXXXXXXXXXXXXXXXXXXXXXX}

\begin{document}

\maketitle

\thispagestyle{plain}
\pagestyle{plain}

\section{Introduction}

The monolithic software architecture, which interweaves all aspects of software logic into a single code base rather than define architecturally separate components, is arguably the most classical way to design software. However, it is difficult to modify specific parts of a monolithic application. While the number of internet global users increased to more than 5 billion in 2024~\cite{internet_usage_2024, internet_usage_2024_2} and with increasing hardware, software, and workload complexity, alternative software architectures that support modularity, adaptability, and scalability have therefore been increasingly adopted. 

In particular, around 2014, Fowler and Lewis~\cite{fowler_microservices} defined and popularized the microservice architecture. It leverages the idea of splitting large monolithic applications into multiple smaller services that interact using lightweight communication schemes. Each service can then scale and be deployed independently from the other parts of the system~\cite{microservice_quality_survey}. The microservice architecture has gotten increasingly more popular in recent years, with over 50\% of the participants in a survey~\cite{microservice_usage} declaring that their organization has adopted the microservice architecture. While there are more modern architectures that try to deal with workload and scalability issues such as serverless~\cite{serverless_usage} and stream processing~\cite{stream_processing_usage}, microservice architecture is still reportedly the most used in the industry~\cite{microservice_usage_icepanel}.

While the microservice architecture has proven its ability to support modern business applications, it also introduces new possible weak points in a system. Microservice architecture can introduce new vulnerabilities associated with the added infrastructure elements~\cite{microservice_inter_service_security} and having the application distributed amongst multiple microservices makes localizing bugs within the system more challenging~\cite{microservice_fault_localization_survey_jacopo}. Some scientific literature surveys have already addressed fault tolerance~\cite{microservice_challenges,microservice_quality_survey} or security~\cite{microservice_security_overview, microservice_security_NIST, microservice_security_review} concerns for microservice architecture but most of them lack analysis on the fault and vulnerability coverage that is introduced by this architecture. Identifying the coverage can be important for risk analysis to evaluate if microservice architecture is appropriate architecture for the application being implemented.

These surveys additionally highlighted a lack of focus on both detection and recovery mechanisms for security attacks~\cite{microservice_security_multivocal_lit_trends} and availability faults~\cite{microservice_availability_tactics}. Problems like these can noticeably degrade business profitability, since according to Pingdom, system downtime can cause businesses to lose from 146k\$ to 450k\$ per hour~\cite{pingdom_downtime_cost}, so having fast detection and recovery procedures for microservice systems is detrimental.

We explore the known faults and vulnerabilities that microservice architecture might suffer from, and the recent scientific literature that addresses them. We emphasize runtime detection and recovery mechanisms instead of offline prevention and mitigation mechanisms to limit its scope. We address the following research questions in the context of microservice architectures:

\begin{enumerate}
    \item What is the state-of-the-art of runtime fault tolerance and security? 
    \item Which runtime faults and vulnerabilities can be detected? 
    \item When and how is it possible to recover from runtime faults and vulnerabilities?
    \item Are there known faults and vulnerabilities that are not addressed by the  literature?
\end{enumerate}

This survey is structured as follows. \hyperref[section: background]{Section~\ref*{section: background}} first presents necessary background knowledge on microservice architectures, dependability and threat modeling. \hyperref[section: related work]{Section~\ref*{section: related work}} addresses and examines related work, including previous surveys, in the field of microservice fault detection and localization. \hyperref[section: methodology]{Section~\ref*{section: methodology}} explains the procedure we followed to obtain the primary literature we discuss. The information we gathered from this literature analysis is split into three sections: meta analysis (\hyperref[section: meta analysis]{Section~\ref*{section: meta analysis}}); fault detection (\hyperref[section: fault detection]{Section~\ref*{section: fault detection}}); and fault recovery (\hyperref[section: fault recovery]{Section~\ref*{section: fault recovery}}). The meta analysis section addresses our first research question, while the fault detection and localization sections respectively address the second and third research questions. \hyperref[section: fault coverage]{Section~\ref*{section: fault coverage}} addresses the fourth research question by analysing the literature and highlighting promising solutions. \hyperref[section: discussion]{Section~\ref*{section: discussion}} discusses and addresses our key findings and the limitations of this survey. Finally, \hyperref[section: conclusion]{Section~\ref*{section: conclusion}} concludes this document.

\section{Related work}
\label{section: related work}

To the best of our knowledge, this survey is the first to review the recent literature on the vulnerabilities and potential faults of microservice architecture, and the proposed solutions to it. 

Several surveys discuss the challenges that microservice architectures face~\cite{microservice_quality_survey, microservice_challenges}. While they give a good overview of the progress of research, they do not focus on fault tolerance. While Li et al.~\cite{microservice_quality_survey} covers common fault tolerance techniques such as circuit breaking, fault or security monitoring, and intrusion recovery, the methods they discuss cover a subset of the threats we consider. Söylemez~\cite{microservice_challenges} follows a similar structure by identifying actively researched challenges in microservices, and finding proposed solutions for them in literature. It includes papers discussing fault tolerance for service discovery, performance monitoring, orchestration platforms, and root cause analysis tools. this survey does not identify gaps in microservice resiliency against faults and threats.

There are also surveys that are more specialized for specific type of fault tolerance mechanisms~\cite{microservice_fault_diagnosis_preprint, microservice_fault_localization_survey_jacopo}. S. Zhan et al.~\cite{microservice_fault_diagnosis_preprint} extensively surveys fault diagnosis and localization techniques presented in the literature since 2003. Soldani and Brogi~\cite{microservice_fault_localization_survey_jacopo} analyze fault diagnosis and localization separately, since methods of fault diagnosis may impact the reliability of the localization methods. 

Surveys that explore security aspects seem to adopt a more similar structure to this survey, where both vulnerabilities and proposed detection or recovery mechanisms are discussed. Multiple papers go over all current literature on MSA security, and map what vulnerabilities and solutions are popular~\cite{microservice_security_multivocal_lit_trends, microservice_security_overview, microservice_inter_service_security}. Chandramouli~\cite{microservice_security_NIST} goes into more detail on how to analyze threats, using a layered approach proposed by Yarygina and Bagge~\cite{microservice_security_layered_model}. These layers are hardware, virtualization, cloud/host, communication, service/app, and orchestration. This helps to identify different kinds of threats that the system may be vulnerable to. Another study addresses current security threats present in the microservice architecture~\cite{microservice_security_review}. Their survey mostly focuses on prevention aspects of security threats rather than recovery mechanisms for security breaches. While these surveys present both the threats and solutions to them, they do not address faults, and some regard the absence of recovery mechanisms in scientific literature ~\cite{microservice_security_review, microservice_security_multivocal_lit_trends, microservice_security_mapping}.

\section{Background}
\label{section: background}

This section gives a primer on microservice architectures and introduces important terms and concepts regarding dependability and threat modeling. 

\subsection{The microservice architecture}
\label{section: micorservices}

A microservice architecture consists of many small and independently deployable services modeled around a business application. It is considered the successor of Service Oriented Architecture (SOA) with clearer attributes and more widely accepted standards. According to Fowler et al.~\cite{fowler_microservices}, there are 9 common characteristics that are shared amongst microservices, most notably componentization via services, decentralized governance, and decentralized data management. Similar characteristics were also proposed by Newman~\cite{newman_microservices}. 

\begin{figure*}[h!]
    \centering
    \includegraphics[width=0.9\textwidth]{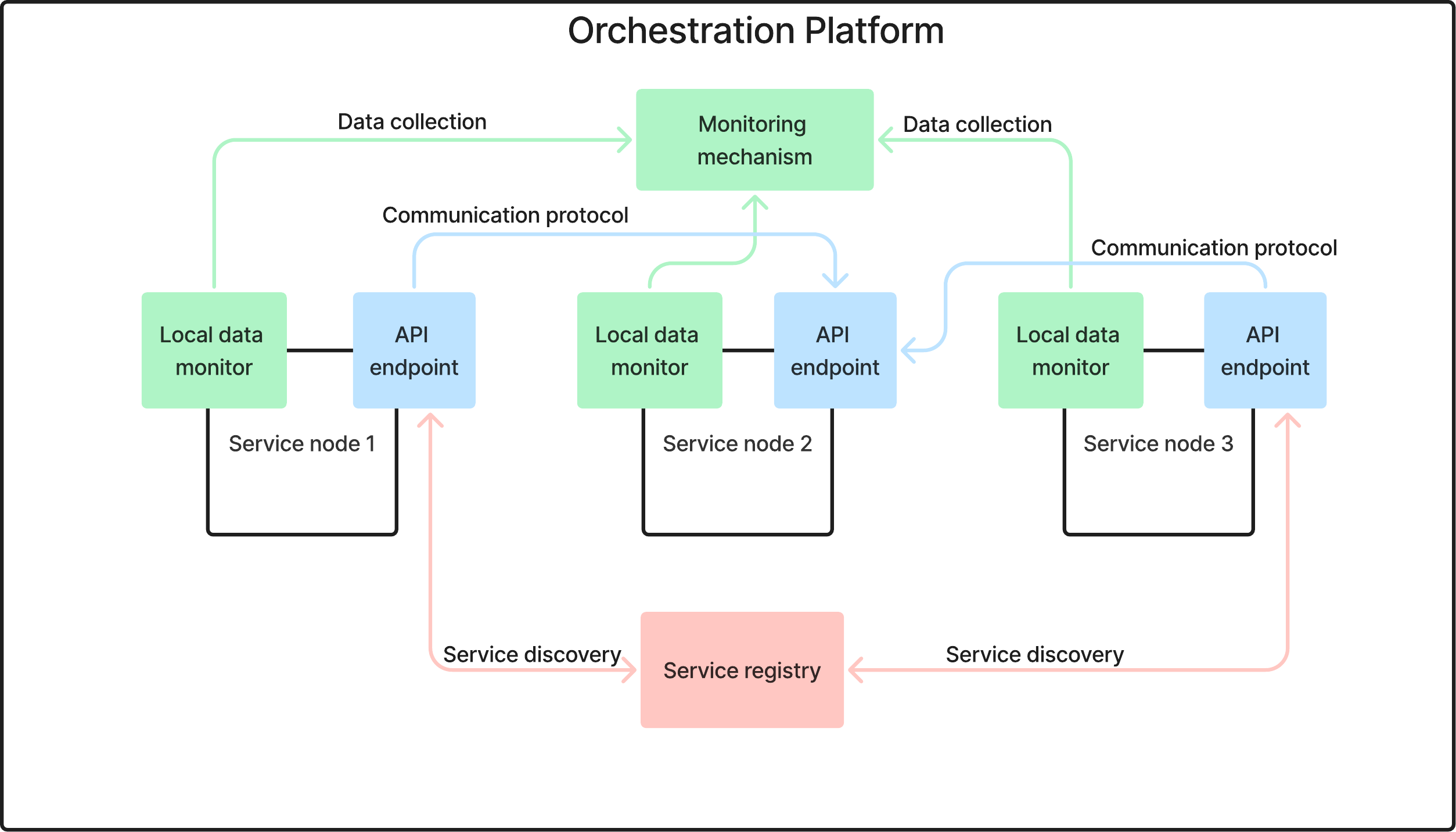}
    \caption{An example of a microservice architecture with a centralized service registry and API endpoints for communication}
    \label{fig: microservice architecture}
\end{figure*}

A visual example of a typical microservice architecture is shown in \autoref{fig: microservice architecture}. It is an orchestration of many smaller applications that provide an overall service to the end users interacting with it. While there is no unifying design of microservice architecture, there are general infrastructure elements that are used that make up the architecture:

\begin{itemize}
    \item \textbf{Microservice nodes} - Independent nodes that serve a specific business logic. These nodes may have multiple replicas to improve application performance or introduce resiliency. These nodes do not share any state and inquire about one another via a communication protocol (detailed below). We consider these nodes to be business logic agnostic for this analysis.
    \item \textbf{Service discovery} - Services may change their location and add additional instances to accommodate user demand. To find active instances of other services, a service registry is used that keeps track of all the nodes and their current location. This infrastructure element can be centralized, which would require another entity in the architecture to facilitate it, or decentralized and embedded into microservice nodes. 
    \item \textbf{Management and deployment platform} - a platform used to manage resources allocated to services, and orchestrate them. This also includes access management between the nodes.
    \item \textbf{Monitoring} - to perform any reaction to the system state, the status of the system has to be known. This can be achieved by monitoring logs, metrics, or any other kind of traces performed by services.
    \item \textbf{Communication protocol} - A lightweight protocol that allows microservice nodes to communicate with one another. Since there is no standard protocol that is used in microservices, we assume that the chosen protocol delivers messages (within time constraints) between nodes without loss, corruption, or creation of new messages. This element also includes components relating to API gateways and endpoints.
\end{itemize}

Microservice applications tend to be large and require many resources to sustain user demand. This makes deployment and resource management complex, usually requiring an orchestration platform or cloud computing platforms to effectively organize the system~\cite{microservice_deplyoment}. There are many tools available that ease this process of resource management such as Kubernetes~\cite{kubernetes} and Docker~\cite{merkel2014docker}, and platforms that allow for deploying on the cloud such as Amazon AWS Lambda or Google Cloud. Each of these technologies come with their own set of possible vulnerabilities or points of failure. To limit the scope of this study, only papers that address a vulnerability in microservice architecture will be addressed, and vulnerabilities present with tools for microservice management or deployment will be omitted.

\subsection{Dependability}

According to the seminal taxonomy of Avižienis et al.~\cite{distributed_system_dependability_taxonomy}, \textbf{dependability} is defined as the ability to avoid service failures that are more frequent and more severe than is acceptable to the users. There are 6 attributes that encapsulate dependability:

\begin{itemize}
    \item \textbf{Availability} - there is sufficient supply of correct services
    \item \textbf{Reliability} - the correct services function as expected
    \item \textbf{Safety} - there is no damage that can be caused to users or the environment
    \item \textbf{Confidentiality} - information is accessible only to sufficiently authorized entities
    \item \textbf{Integrity} - provided information is not accidentally or maliciously modified.
    \item \textbf{Maintainability} - system can be modified to suit user needs.
\end{itemize}

Each attribute encompasses a large field of research within the microservice architecture. In this survey, we will concentrate on availability, reliability, confidentiality, and integrity attributes, since they are directly related to the research questions. The concept of avoiding system failures in the presence of faults is called \textbf{fault tolerance}~\cite{distributed_system_dependability_taxonomy}.

The terms fault, vulnerability, error, and failure tend to be used interchangeably when addressing dependability in microservice architecture. Avižienis et al.~\cite{distributed_system_dependability_taxonomy}  present a useful fault-error-failure model that gives a specific meaning to each of the terms when modeling failure in distributed systems. We additionally use the term \textbf{vulnerability} to define specific types of faults that allow users to intentionally or unintentionally harm the system and cause other faults.

A \textbf{failure} occurs when a system fails to provide a service according to its specification. These failures can be classified into 3 major groups: content, timing, or a combination of content or timing. For each of these failure groups, different kinds of errors can be the causes of this malfunction. \textbf{Errors} are states within services that may lead to a service failure. Content failure can be caused by accessing unauthorized information or incorrect information errors. Timing failures can either relate to late or early message arrival or delivery errors. And finally, the combination of these failures occurs when the services either halt/crash and produce no output, or produce erratic output.

An error is caused by an underlying fault. \textbf{Faults} are the lowest level system protocol deviations. They can be again subdivided into 3 general categories: development, physical, and interaction faults. Development faults are all the faults that are caused during development, such as logic bugs or misconfiguration. These faults will be disregarded in this study since they relate more to offline fault mitigation tactics, rather than runtime fault detection and recovery. Physical faults relate to any kind of hardware component failure, while interaction faults relate to all the faults that can be caused by interactions with the application.

\subsection{Threat modeling}
\label{section: threat}

According to The Open Worldwide Application Security Project (OWASP), threat modeling is the process of identifying threats and defining countermeasures to mitigate and prevent the effects of each threat~\cite{owasp_threat_modeling}. This encapsulates both malicious (performed intentionally by a 3rd party member) or incidental threats (such as hardware failure). OWASP additionally proposes a simple 4-step framework for analyzing system vulnerabilities~\cite{owasp_threat_modeling}:

\begin{enumerate}
    \item Assessment scope
    \item Identify what can go wrong
    \item Identify countermeasures
    \item Assess your work
\end{enumerate}

Assessment scope refers to the underlying system that will be examined. In this survey, each infrastructure element that is presented in \autoref{section: micorservices} will be used for this analysis. To identify what can go wrong, faults and vulnerabilities will be examined from the perspective of each of these components. To mitigate these faults, countermeasures from available literature will be used to mitigate them. Lastly, the presented methods will be analyzed in their quality in addressing system vulnerabilities, and what gaps there still exist within the literature.

There are other popular threat models such as STRIDE and PASTA~\cite{microservice_security_review}. PASTA stands for Process for Attack Simulation and Threat Analysis. PASTA threat model is more intertwined with the business logic and underlying technologies of the application, hence does not fit very well with our generalized model. STRIDE on the other hand stands for Spoofing, Tampering, Repudiation, Information disclosure, Denial of service and Elevation of privilege. Unlike PASTA, this model is more generic and helps to identify common threats, hence will be used to complement the OWASP threat model.

\subsection{Microservice faults}
\label{section: faults}

By trying to identify all possible faults in a system, appropriate fault tolerance mechanisms can be introduced to avoid system failure.

We identified two studies that respectively try to summarize observed faults~\cite{microservice_fault_taxonomy} and security vulnerabilities~\cite{microservice_sercurity_taxonomy} within microservice architecture. F. Silva et al.~\cite{microservice_fault_taxonomy} construct a microservice fault catalog according to ISO 25000 NFRs quality standards. R. K. Jayalath et al.~\cite{microservice_sercurity_taxonomy} examined 62 studies to find inherent vulnerabilities of microservice architecture and categorize them according to their distinctive features. We use both of these works as the basis for identifying faults faced within the microservice architecture. There are 117 described faults and 126 identified vulnerabilities in each paper respectively, so for readability and relevancy purposes, some faults and vulnerabilities are discarded and the rest are generalized into categories. Categories are grouped into either performance \textbf{(P)}, architecture \textbf{(A)}, component \textbf{(C)}, or security \textbf{(S)} related groups. This is done to make referencing more descriptive and provide a better overview of covered domains. Identified categories are as follows:

\begin{itemize}
    \item \textbf{Memory performance fault (P1)} - this category relates to all faults that cause the main memory of the application to perform slower than expected. This includes memory leaks, memory allocation faults, or memory anomalies.
    \item \textbf{CPU performance fault (P2)} - this encapsulates all the faults that relate to reducing the expected CPU performance, such as CPU hogging, CPU resource allocation faults, or CPU anomaly.
    \item \textbf{Disk performance fault (P3)} - this addresses faults that cause disk I/O to underperform. This includes I/O errors, disk usage spikes, or disk anomalies.
    \item \textbf{Deadlock (P4)} - Deadlock occurs when the system is unable to progress due to each process waiting for resources to be released. This category also contains resource starvation, where a process is unable to obtain the resources needed for completing the computation.
    \item \textbf{Process crash (P5)} - this relates to any faults that cause the entire process to crash.
    \item \textbf{Message delay (A1)} - this relates to when a microservice does not receive the necessary response from another service in time to continue computation.
    \item \textbf{Configuration fault (A2)} - any form of misconfiguration done within the service that causes it to deviate from the protocol. This also includes version incompatibility faults.
    \item \textbf{User or service interaction fault (A3)} - this relates to any authorized action performed by a user or a service that still results in the system misbehaving. This includes business-level bugs, database failures, or other integrated tool failures.
    \item \textbf{Insecure confidential data (S1)} - these faults relate to any parts of the business application that can be accessed by an unauthorized party. 
    \item \textbf{Service hijacking (S2)} - this relates to giving unauthorized users permission to control any part of the system.

\end{itemize}

STRIDE~\cite{microservice_security_review}, OWASP Top Ten~\cite{owasp_top_10}, and the overview made by A. Hannousse and S. Yahiouche~\cite{microservice_security_overview} were used to complement this list, such as:

\begin{itemize}
    \item \textbf{Service registry corruption (C1)} - if the service registry crashes or becomes corrupt, this can cause microservices to not be able to establish communication channels. This is especially an issue for centralized service registry implementations.
    \item \textbf{Service registry hijacking (C2)} - malicious actor could modify the service registry, or impersonate one. This includes replay attacks to extract confidential information.
    \item \textbf{Monitor data corruption (C3)} - monitoring data may be corrupted or missing for certain periods of time. This can drastically affect decision-making for recovery algorithms.
    \item \textbf{IoT device failures (A4)} - there are many different IoT devices used for real-time measurements or actions, such as sensors and actuators. These devices can give incorrect outputs or behaviors, or not work at all.
    \item \textbf{External denial of service (S3)} - this encapsulates situations where a malicious actor denies service to other users, usually by flooding the service with too many requests.
    \item \textbf{Internal denial of service (S4)} - this occurs when another microservice apart of the system enacts the denial of service of some part of the system because of malfunction or malicious behavior.
    \item \textbf{Malicious injection (S5)} - this involved a malicious party inserting damaging logic within the service infrastructure, that can either cause failures or leak information. This encapsulates Cross-Site Request Forgery and code injection.

\end{itemize}

In \autoref{table:servicefailure}, the faults are categorized within the fault-error-failure model. Additionally, related infrastructure elements that suffer from the fault are added at the end of the table. Early message arrival has no faults attributed to it since the overwhelming majority of literature tries to reduce the messaging time to as low as possible.

\renewcommand{\arraystretch}{1.3}
\begin{table*}[h!]
\centering

\newcolumntype{M}{>{\begin{varwidth}{4cm}}l<{\end{varwidth}}} 

\caption{Fault-error-failure model of microservice architectures}
\label{table:servicefailure}
\begin{tabular}{|l|l|l|p{7cm}|}
\hline
\textbf{Service Failures} & \textbf{Errors} & \textbf{Faults} & \textbf{Infrastructure Elements} \\ \hline
\multirow{2}{*}{Content} & Unauthorized information & Insecure confidential data & Microservice node; Management platform; Communication protocol \\ \cline{2-4}
& Incorrect information & Disk performance fault & Microservice node \\ \cline{3-4}
& & IoT device failure  & Microservice node  \\
 \hline

\multirow{4}{*}{Timing} & Early message arrival & - & - \\ \cline{2-4}
 &\multirow{4}{*}{Late message arrival}  & CPU performance fault & Microservice node \\ \cline{3-4}
 & & Memory performance fault & Microservice node \\ \cline{3-4}
 & & Disk performance fault & Microservice node \\ \cline{3-4}
 & & Message delay & Microservice node; Service discovery; Management platform; Communication protocol \\ \hline

\multirow{9}{*}{Content \& Timing} & \multirow{5}{*}{Halt/crash}  & Deadlock & Microservice node; Service discovery \\ \cline{3-4}
 & & Process crash & Microservice node \\ \cline{3-4}
 & & Service registry corruption & Service discovery \\ \cline{3-4}
 & & Monitor data corruption & Monitoring mechanism \\ \cline{3-4}
 & & Internal denial of service (DoS) & Microservice node \\ \cline{3-4}
 & & External denial of service (DoS) & Management platform; Communication protocol \\ \cline{2-4}
 & \multirow{4}{*}{Erratic response}  & User interaction fault & Microservice node \\ \cline{3-4}
 & & Malicious injection & Microservice node\\ \cline{3-4}
 & & Service hijacking & Microservice node; Communication protocol \\ \cline{3-4}
 & & Configuration fault & Microservice node; Management platform \\ \cline{3-4}
 & & Service registry hijacking & Service Registry; Communication protocol \\ \hline

\end{tabular}
\end{table*}

\section{Methodology}
\label{section: methodology}

To ensure the quality and objectiveness of this survey, guidelines suggested by H. Zhang et al.~\cite{systematic_review_quasi} and Snyder~\cite{systematic_review_generic} were partially followed. A systematic approach for conducting the survey was chosen because it has the potential to minimize reliance on the researcher's background knowledge and increase the relevancy of the found literature~\cite{systematic_review_quasi}. However, due to the time and scope limitations of this survey, some systematic procedures were not followed such as constructing the 'quasi-gold standard'~\cite {systematic_review_quasi} or using objective methods of keyword extraction for search queries. While this may introduce more bias and dismiss some relevant literature works, this impact should be minimized by using snowballing to potentially find papers that were not found using the search query.

\begin{figure}
    \centering
    \includegraphics[width=0.65\linewidth]{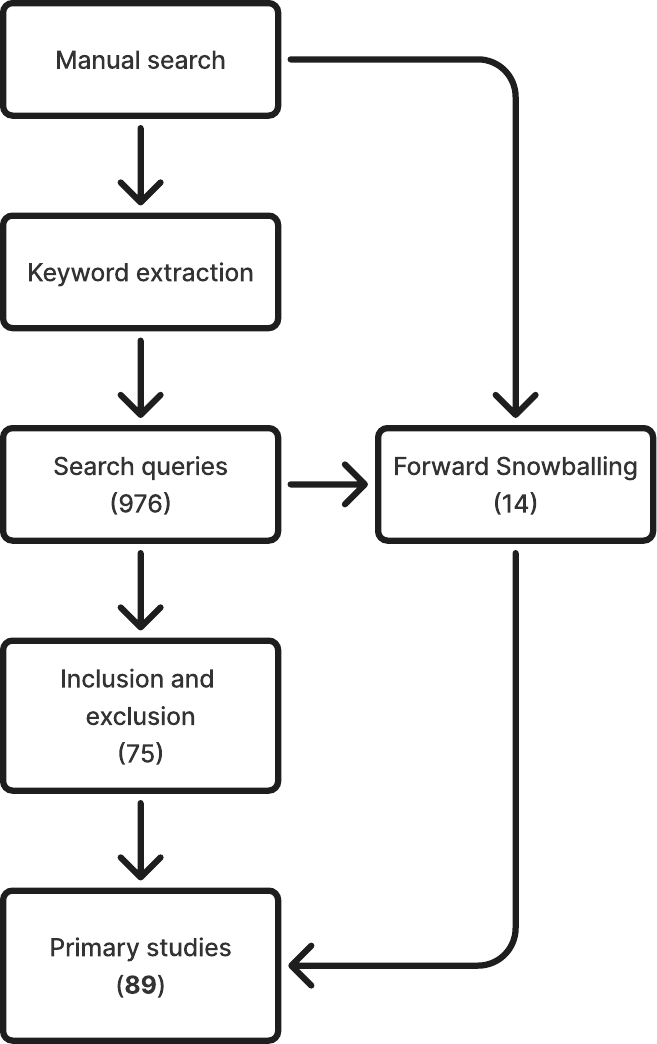}
    \caption{Method followed to identify primary literature}
    \label{fig: method}
\end{figure}

First, a superficial screening of the research space was done by using the \textit{microservice fault tolerance survey} and \textit{microservice security survey} queries on Google Scholar. These studies were mainly used to extract important keywords and verify the identified research gap. Using the keywords and the proposed research question, a search query was constructed. To limit the scope and ensure the quality of the examined literature, clear inclusion and exclusion criteria are defined. Finally, to ensure better coverage of relevant works, forward snowballing is used to find papers that are related to the research question but were omitted by the search query. A simple overview of our approach is showcased in \autoref{fig: method}

\subsection{Search queries}

IEEE Xplore and ACM Digital Library were chosen as the main platforms to conduct the literature search. The main reason for choosing these libraries is their relevance to the computer science research field, the large amount of available literature they reference, their support for complex queries, and their popularity among other systematic literature surveys~\cite{systematic_review_quasi}. Later on, to increase the literature coverage, Web of Science was utilized. Web of Science is a bibliographic database that allows to query multiple scientific journals and databases for scientific research. This allowed to include papers from a more diverse set of digital research libraries.
To construct the queries, important keywords were identified among studies mentioned in \autoref{section: related work}. To further refine it, we adjusted the query to include or omit related terms. The final queries can be found in \autoref{section: queries}.

The conjunctive normal form was used to construct the queries because some platforms did not support the complex mixing of boolean operators. To reduce irrelevant literature found, only the metadata of the papers was used to match the search query. For each platform, what is considered metadata differs, but generally, metadata includes the title, keywords, and abstract, while always excluding the full text. Wildcard character is used frequently to include different conjugations of words. 

To limit the scope, papers written in the past five years were collected, from January 2019 until 2024 October (the time of writing this survey). The total number of studies obtained was 976, where 685 were obtained from IEEE Xplore, 104 from the ACM Digital Library, and 187 from Web of Science. 

\subsection{Inclusion/exclusion criteria}

To systematically filter irrelevant or not peer-reviewed papers, explicit inclusion and exclusion criteria were defined (see \autoref{table:inclusioncrit}). These explicit criteria help to systematically include and exclude papers for this survey. While literature reviews and surveys were excluded, notable ones are addressed in \autoref{section: related work} and used for snowballing.

Initially, the inclusion/exclusion criteria were applied by reading the title and abstract of the paper. In this way, some false negatives may have been discarded from this review, which we discuss in \autoref{section: discussion}. After reading the full text of the paper, some false positives were additionally discarded. After filtering the results, \textbf{75} papers were kept.

\begin{table}[t]
\centering
\caption{Inclusion and Exclusion Criteria for Paper Selection}
\label{table:inclusioncrit}
\begin{tabular}{@{}p{4cm}p{4cm}@{}}

\toprule
\textbf{Inclusion Criteria} & \textbf{Exclusion Criteria} \\ \midrule
1. The paper is within the field of microservice architecture. & 1. Topics related to fault tolerance or security are not the primary focus of the paper \\

2. The paper proposes a novel solution or improvement to runtime fault discovery or recovery.  & 2. Secondary studies, such as literature reviews or surveys. \\

3. The paper is published or a part of a scientific journal or a book.                            & 3. The proposed solution is only applicable to specific business logic or applications. \\

4. The paper is accessible freely by a TU Delft University member. & 4. The paper only addresses offline fault mitigation techniques (code analysis, testing, etc.). \\

\bottomrule
\end{tabular}
\end{table}

\subsection{Forward snowballing}

Forward snowballing is a technique where a set of original studies are found using search queries, and the subsequent papers that cite them are examined. Backward snowballing looks instead at papers cited by the original selected studies. Snowballing is helpful to identify all relevant studies on a given topic, and in particular those that may have been excluded by the original search queries.

In this survey, we only use forward snowballing to find the most recent literature that has not been found by the search query. We use papers from the related work section~\cite{microservice_quality_survey, microservice_challenges, microservice_fault_diagnosis_preprint, microservice_fault_localization_survey_jacopo, microservice_security_overview, microservice_security_multivocal_lit_trends, microservice_security_NIST, microservice_security_review, microservice_inter_service_security,
microservice_security_layered_model} and influential papers found by the search query~\cite{60_automap, 56_semi_supervised_vae, 49_latent_error, 33_microrca, 72_trace_comparison} that have over 20+ citations. In total, \textbf{14} new papers were added to the literature pool.

\subsection{Data extraction}

For this survey, not all details of the papers are relevant for analysis, hence only certain parts are extracted to answer the research questions. These data items are described below:

\begin{itemize}
    \item \textbf{Year of publication} - used for meta-analysis, to see how many papers are published on this topic per year, indicating either growth, stagnation, or decline of this research field.
    \item \textbf{Addressed faults, anomalies, or threats} - helps classify the system faults that a paper addresses and how they relate to the threat model.
    \item \textbf{Solution} - addresses the method the paper uses to solve the proposed system failure(s). This will both be used to group solutions into similar categories and explain the proposed implementation in more detail.
    \item \textbf{Evaluation techniques} - allows us to compare different solutions and their efficiency in providing fault tolerance.
\end{itemize}

In the following sections, we will use this extracted data to make observations about the reviewed literature.

\section{Meta analysis}
\label{section: meta analysis}

In this section, we perform a meta-analysis to identify research trends. First, we examine the growth of literature per year to see the popularity of the field. Then we examine the most important keywords that appear in the surveyed literature. 

\subsection{Publications per year}
\begin{figure*}[h]
    \centering
    \includegraphics[width=0.7\linewidth]{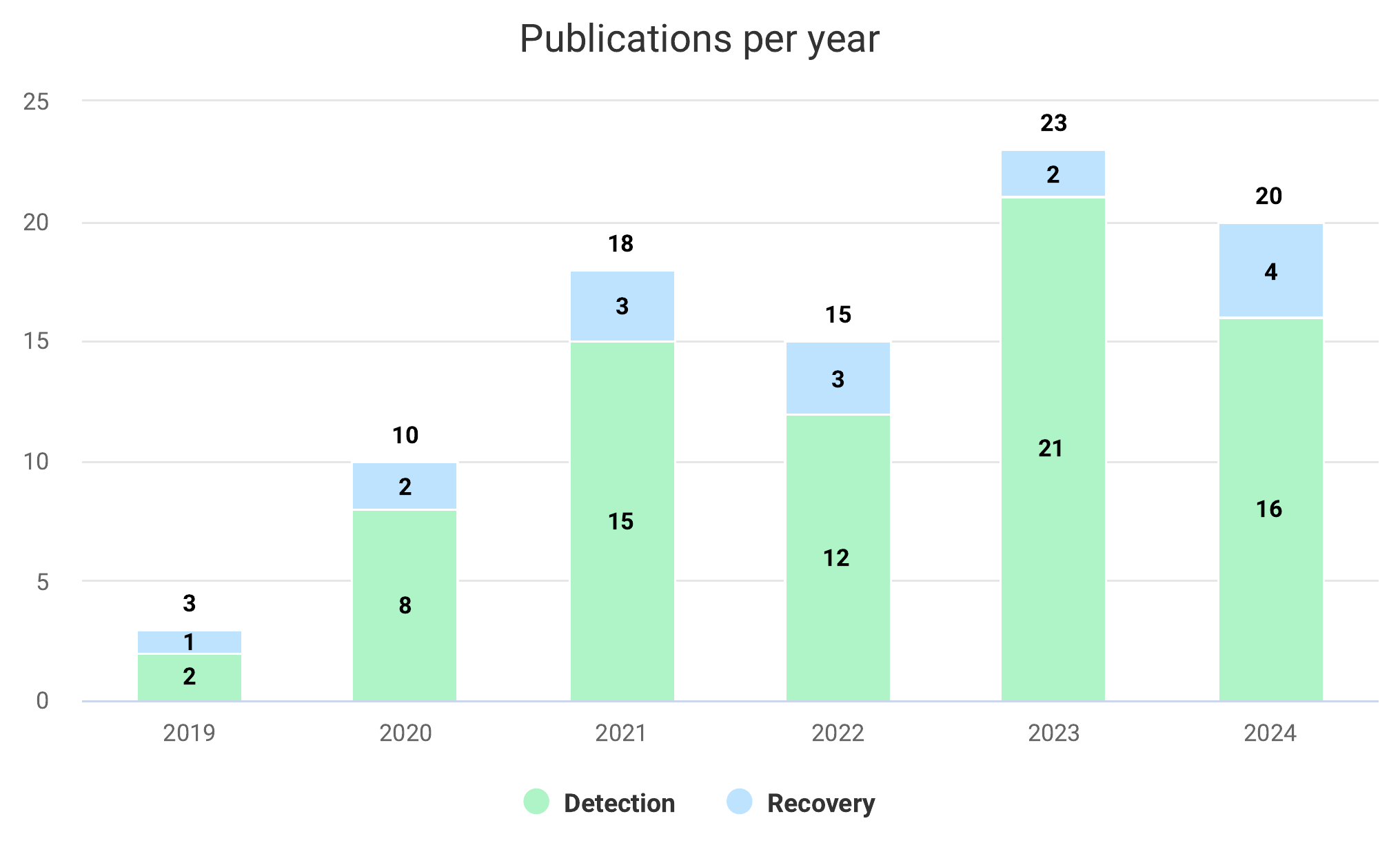}
    \caption{Number of selected publications per year}
    \label{fig: publications per year}
\end{figure*}

To estimate the research interest in the topic, we look at the number of publications per year. The trend is shown in \autoref{fig: publications per year}. It can be observed that there is a steady increase in the number of papers each year, with an average growth rate of 87\% from the year 2019 to 2023. Note that this survey included papers until October 2024, meaning that more papers may have been published since.

It seems that research is significantly more focused on detection methods for faults, rather than having automatic ways to recover from them. The ratio between recovery to detection related surveyed literature is almost 1:5. This shows that the research field is more focused on finding novel ways of detecting faults within the system, rather than researching ways how to recover from said faults.

\subsection{Keyword}

To identify frequent and important keywords within the research community, we use the approach and visualization methods proposed by Versluis and Iosup~\cite{laurens_paper}. Term Frequency–Inverse Data Frequency (TF-IDF) allows for highlighting keywords within a research domain by analyzing the frequency of reoccurring words within each surveyed paper, and discarding keywords that appear frequently within all literature that are not specific to this community.

\begin{figure*}[h]
    \centering
    \includegraphics[width=1\linewidth]{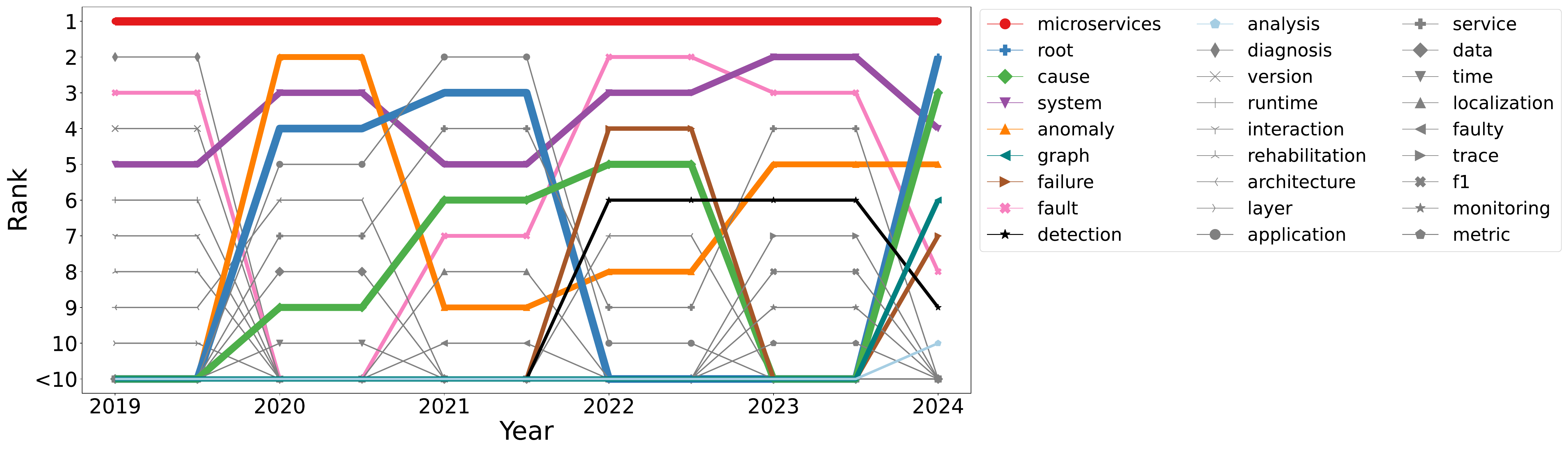}
    \caption{Evolution of the ranking of the 10 most important keywords ranking depending on the year using TF-IDF}
    \label{fig: keywords}
\end{figure*}

The changes of most used keywords within the research community over the span of 6 years can be observed in~\autoref{fig: keywords}. As anticipated, the most important keyword identified by TF-IDF is microservice, since it is the core infrastructure the solutions are built around. Most interesting observation is that there seems to be no consensus on which keyword is used to identify misbehavior within the system, with keywords "anomaly", "failure", and "fault", all fluctuating in popularity throughout the years. Many keywords that relate to fault detection are popular, such as "root", "cause", "detection", and "diagnosis", "runtime", "localization" and "monitoring". Interestingly, only "trace" and "metric" modalities appear in the top 10, but only for a single year. Finally, the keyword that may indicate a new trend in this community that has emerged is "graph", suggesting that algorithms that utilize system graphs may become more prevalent in the upcoming years.

In the following sections, a more individual analysis of each paper will be performed, and relevant data will be summarized in tables.

\section{Fault detection}
\label{section: fault detection}

The majority of the fault detection literature is split into two main groups: anomaly detection and fault localization (also referred to as root cause analysis). Anomalies are any form of deviation from normal behaviour. This definition includes both faults that were defined in \autoref{section: faults}, and other deviations that may not materialize into a fault. This additionally gives an opportunity for fault prediction, where abnormalities are detected early and handled before inducing a fault~\cite{9_fsps_fault_prediction, 10_indicator_prediciton, 12_automated_monitoring_lstm, 62_SuanMing}. Fault localization tackles the problem of finding the service or event that induced the fault within the system. As the number of microservice nodes grows, it becomes more difficult to identify the root cause of a fault. 

Anomaly detection and fault localization methods both rely on runtime data. There are 3 main types of data that can be used to detect anomalies within a system:

\begin{itemize}
    \item \textbf{Logs (L)} - records collected by services that show the events they were sujected to. These logs can also be combined into application level logs that give a higher level overview of events within the system.
    \item \textbf{Traces (T)} - records of data requests as they flow through the application. This usually requires instrumentation to be applied to services, so that more granular analysis of node relationships can be performed.
    \item \textbf{Metrics (M)} - records of component or application measurements. These can be used to observe how the performance, availability and reliability vary during runtime.
\end{itemize}

Some papers used additional information such as kernel traces, microservice topology, request metadata, configuration data, or user information for their detection mechanism. We classify these as \textbf{miscellaneous (Mi)} data types. Some implementations use a single type of data for anomaly detection (single-modal), while others attempt to combine types of data to use all possible sources of information for more accurate localization (multi-modal).

To assist in understanding and recovering from a fault, some solutions also include fault classification. This usually is separated from fault localization, since it only includes locating the fault, without necessarily giving more details about its type.

To give a simple overview of analyzed papers, their attributes and fault coverage are summarized in \autoref{table:anomalytablesinglemodal} for single-modal approaches and in \autoref{table:anomalytablemulti-modal} for multi-modal approaches. The tables showcase what type of data is utilized within a paper, whether the paper performs detection, localization, or classification, and what the kinds of faults they cover (defined in \autoref{section: faults}).

The next sections give a brief explanation of the methods used in these papers.

\begin{table*}[!h]
	\caption{Anomaly detection and fault localization solutions (part 1)  }
        \label{table:anomalytablesinglemodal}
	\begin{center}
		\rowcolors{3}{gray!15}{white}
		\resizebox{\textwidth}{!}{
			\begin{tabular}{p{12mm} p{2mm}p{2mm}p{2mm}p{2mm}p{2mm}p{2mm}p{2mm}p{2mm}p{2mm}p{2mm}p{2mm}p{2mm}p{2mm}p{2mm}p{2mm}p{2mm}p{2mm}p{2mm}p{2mm}p{2mm}p{8mm}}
				\toprule
												
				\\

				\textit{Papers} & 
				\rotatebox{60}{\textit{Modality}} & 
				\rotatebox{60}{\textit{Detection}} & 
				\rotatebox{60}{\textit{Localization}} & 
				\rotatebox{60}{\textit{Classification}} & 
				\rotatebox{60}{\textit{Memory (P1)}} &
				\rotatebox{60}{\textit{CPU (P2)}} & 
				\rotatebox{60}{\textit{Disk (P3)}} & 
				\rotatebox{60}{\textit{Deadlock (P4)}} &
				\rotatebox{60}{\textit{Process crash (P5)}}  &

				\rotatebox{60}{\textit{Message delay (A1)}} & 
				\rotatebox{60}{\textit{Configuration (A2)}} & 
				\rotatebox{60}{\textit{User interaction (A3)}}  & 
				\rotatebox{60}{\textit{IoT device (A4)}} &
												
				\rotatebox{60}{\textit{Message delay (C1)}} & 
				\rotatebox{60}{\textit{Configuration (C2)}} & 
				\rotatebox{60}{\textit{User interaction (C3)}}  &

				\rotatebox{60}{\textit{Insecure data (S1)}} &
				\rotatebox{60}{\textit{Service hijacking (S2)}} &
				\rotatebox{60}{\textit{External DoS (S3)}} &
				\rotatebox{60}{\textit{Internal DoS (S4)}} &
				\rotatebox{60}{\textit{Injection (S5)}} 
				\\   
				\midrule 

				\cite{9_fsps_fault_prediction}                                                    & M & X & X & \multicolumn{1}{p{2mm}|}{X} & X & X & X & X & \multicolumn{1}{p{2mm}|}{X} & X &   &   & \multicolumn{1}{p{2mm}|}{}  &   &   & \multicolumn{1}{p{2mm}|}{}  &   &   &   &                                                 &   \\
				\cite{10_indicator_prediciton}                                                    & M & X &   & \multicolumn{1}{p{2mm}|}{}  &   &   &   &   & \multicolumn{1}{p{2mm}|}{}  &   &   &   & \multicolumn{1}{p{2mm}|}{}  &   &   & \multicolumn{1}{p{2mm}|}{}  &   &   &   &                                                 &   \\
				\cite{11_anomaly_detection_transformer}                                           & M & X & X & \multicolumn{1}{p{2mm}|}{}  & X & X & X &   & \multicolumn{1}{p{2mm}|}{}  &   &   &   & \multicolumn{1}{p{2mm}|}{}  &   &   & \multicolumn{1}{p{2mm}|}{}  &   &   &   &                                                 &   \\
				\cite{22_fault_localization_impact_graph_impacttracer}                            & M & X & X & \multicolumn{1}{p{2mm}|}{}  & X & X & X &   & \multicolumn{1}{p{2mm}|}{X} & X &   & X & \multicolumn{1}{p{2mm}|}{}  &   &   &\multicolumn{1}{p{2mm}|}{}   &   &   &   &                                                 &   \\
				\cite{31_flowrca}\cite{32_iot_root_cause}\cite{29_microcause_causality_inference} & M & X & X & \multicolumn{1}{p{2mm}|}{X} & X & X & X &   & \multicolumn{1}{p{2mm}|}{X} & X &   &   & \multicolumn{1}{p{2mm}|}{}  &   &   & \multicolumn{1}{p{2mm}|}{}  &   &   &   &                                                 &   \\
												
				\cite{36_dycause}                                                                 & M & X & X & \multicolumn{1}{p{2mm}|}{}  & X & X & X &   & \multicolumn{1}{p{2mm}|}{X} & X &   &   & \multicolumn{1}{p{2mm}|}{}  &   &   & \multicolumn{1}{p{2mm}|}{}  &   &   &   &                                                 &   \\
				\cite{41_neural_multiclassification}\cite{42_tls_wgan_gp}\cite{64_asfc}           & M & X & X & \multicolumn{1}{p{2mm}|}{X} & X & X &   &   & \multicolumn{1}{p{2mm}|}{}  & X &   &   & \multicolumn{1}{p{2mm}|}{}  &   &   & \multicolumn{1}{p{2mm}|}{}  &   &   &   &                                                 &   \\
				\cite{43_component_analysis}\cite{56_semi_supervised_vae}                         & M & X &   & \multicolumn{1}{p{2mm}|}{}  & X & X &   &   & \multicolumn{1}{p{2mm}|}{}  & X &   &   & \multicolumn{1}{p{2mm}|}{}  &   &   & \multicolumn{1}{p{2mm}|}{}  &   &   &   &                                                 &   \\

				\cite{47_topo_rca}\cite{50_latentscope}                                           & M & X & X & \multicolumn{1}{p{2mm}|}{}  & X & X &   &   & \multicolumn{1}{p{2mm}|}{}  & X &   &   & \multicolumn{1}{p{2mm}|}{}  &   &   & \multicolumn{1}{p{2mm}|}{}  &   &   &   &                                                 &   \\
												
				\cite{53_tritium}                                                                 & M & X & X & \multicolumn{1}{p{2mm}|}{}  &   &   &   &   & \multicolumn{1}{p{2mm}|}{}  &   &   & X & \multicolumn{1}{p{2mm}|}{}  &   &   & \multicolumn{1}{p{2mm}|}{}  &   &   &   &                                                 &   \\
				\cite{55_baro}                                                                    & M & X & X & \multicolumn{1}{p{2mm}|}{}  &   & X &   &   & \multicolumn{1}{p{2mm}|}{}  & X &   &   & \multicolumn{1}{p{2mm}|}{}  &   &   & \multicolumn{1}{p{2mm}|}{}  &   &   &   &                                                 &   \\
												
				\cite{60_automap}                                                                 & M & X & X & \multicolumn{1}{p{2mm}|}{}  & X & X & X &   & \multicolumn{1}{p{2mm}|}{}  & X &   &   & \multicolumn{1}{p{2mm}|}{}  &   &   & \multicolumn{1}{p{2mm}|}{}  &   &   & X &                                                 &   \\
												
				\cite{69_mlp}                                                                     & M & X &   & \multicolumn{1}{p{2mm}|}{}  & X & X & X &   & \multicolumn{1}{p{2mm}|}{}  & X &   &   & \multicolumn{1}{p{2mm}|}{}  &   &   & \multicolumn{1}{p{2mm}|}{}  &   &   &   &                                                 &   \\
				\cite{70_causal_discovery}                                                        & M & X & X & \multicolumn{1}{p{2mm}|}{}  & X & X &   &   & \multicolumn{1}{p{2mm}|}{}  &   &   &   & \multicolumn{1}{p{2mm}|}{}  &   &   & \multicolumn{1}{p{2mm}|}{}  &   &   &   &                                                 &   \\
				\cite{84_causalrca}                                                               & M &   & X & \multicolumn{1}{p{2mm}|}{X} & X & X & X & X & \multicolumn{1}{p{2mm}|}{}  &   &   &   & \multicolumn{1}{p{2mm}|}{}  &   &   & \multicolumn{1}{p{2mm}|}{}  &   &   &   &                                                 &   \\
				\cite{87_microdiag}                                                               & M & X & X & \multicolumn{1}{p{2mm}|}{X} & X & X &   &   & \multicolumn{1}{p{2mm}|}{}  &   &   &   & \multicolumn{1}{p{2mm}|}{}  &   &   & \multicolumn{1}{p{2mm}|}{}  &   &   &   &                                                 &   \\

				\cite{16_fault_localization_service_dependency_based}                             & M & X & X & \multicolumn{1}{p{2mm}|}{X} &   & X &   &   & \multicolumn{1}{p{2mm}|}{}  & X & X & X & \multicolumn{1}{p{2mm}|}{}  &   &   & \multicolumn{1}{p{2mm}|}{}  &   &   &   &                                                 &   \\
												
				\cite{17_fault_localization_continuation_of_16}                                   & M & X &   & \multicolumn{1}{p{2mm}|}{X} &   & X &   &   & \multicolumn{1}{p{2mm}|}{}  & X & X & X & \multicolumn{1}{p{2mm}|}{}  &   &   & \multicolumn{1}{p{2mm}|}{}  &   &   &   &                                                 &   \\
												
				\midrule
												
				\cite{37_lograg}                                                                  & L & X &   & \multicolumn{1}{p{2mm}|}{}  & X & X & X &   & \multicolumn{1}{p{2mm}|}{X} & X &   &   & \multicolumn{1}{p{2mm}|}{}  &   &   & \multicolumn{1}{p{2mm}|}{}  &   &   &   &                                                 &   \\
				\cite{40_causal_slo}                                                              & L & X & X & \multicolumn{1}{p{2mm}|}{}  & X & X & X &   & \multicolumn{1}{p{2mm}|}{X} & X &   &   & \multicolumn{1}{p{2mm}|}{}  &   &   & \multicolumn{1}{p{2mm}|}{}  &   &   &   &                                                 &   \\
				\cite{67_golden_signals}                                                          & L & X & X & \multicolumn{1}{p{2mm}|}{}  &   &   &   &   & \multicolumn{1}{p{2mm}|}{}  & X &   &   & \multicolumn{1}{p{2mm}|}{}  &   &   & \multicolumn{1}{p{2mm}|}{}  &   &   &   &                                                 &   \\
				\cite{74_logattention}                                                            & L & X &   & \multicolumn{1}{p{2mm}|}{}  & X & X & X &   & \multicolumn{1}{p{2mm}|}{}  & X &   &   & \multicolumn{1}{p{2mm}|}{}  &   &   & \multicolumn{1}{p{2mm}|}{}  &   &   &   &                                                 &   \\
												
				\midrule
												
				\cite{46_self_supervised}                                                         & T & X &   & \multicolumn{1}{p{2mm}|}{}  & X & X &   &   & \multicolumn{1}{p{2mm}|}{}  & X &   &   & \multicolumn{1}{p{2mm}|}{}  &   &   & \multicolumn{1}{p{2mm}|}{}  &   &   &   &                                                 &   \\
				\cite{76_informer}                                                                & T & X &   & \multicolumn{1}{p{2mm}|}{}  &   &   &   &   & \multicolumn{1}{p{2mm}|}{}  &   &   & X & \multicolumn{1}{p{2mm}|}{}  &   &   &\multicolumn{1}{p{2mm}|}{}   &   &   &   &                                                 & X \\
				\cite{77_workflow_aware}                                                          & T & X & X & \multicolumn{1}{p{2mm}|}{}  &   & X &   &   & \multicolumn{1}{p{2mm}|}{}  & X & X &   & \multicolumn{1}{p{2mm}|}{}  &   &   & \multicolumn{1}{p{2mm}|}{}  &   &   &   &                                                 &   \\
				\cite{90_micronet}                                                                & T & X & X & \multicolumn{1}{p{2mm}|}{}  &   &   &   &   & \multicolumn{1}{p{2mm}|}{}  & X &   &   & \multicolumn{1}{p{2mm}|}{}  &   &   & \multicolumn{1}{p{2mm}|}{}  &   &   &   &                                                 &   \\
												
				\cite{54_groupwise}                                                               & T & X &   & \multicolumn{1}{p{2mm}|}{}  &   & X &   &   & \multicolumn{1}{p{2mm}|}{}  & X &   &   & \multicolumn{1}{p{2mm}|}{}  &   &   & \multicolumn{1}{p{2mm}|}{}  &   &   &   &                                                 &   \\

				\midrule
												
				\cite{1_IoT_hybrid_FT}                                                            & O & X & X & \multicolumn{1}{p{2mm}|}{X} &   &   &   &   & \multicolumn{1}{p{2mm}|}{}  &   &   &   & \multicolumn{1}{p{2mm}|}{X} &   &   & \multicolumn{1}{p{2mm}|}{}  &   &   &   &                                                 &   \\
												
				\cite{6_fault_tolerance_telerehab_iot}                                            & O & X & X & \multicolumn{1}{p{2mm}|}{}  &   &   &   &   & \multicolumn{1}{p{2mm}|}{}  &   &   &   & \multicolumn{1}{p{2mm}|}{X} &   &   & \multicolumn{1}{p{2mm}|}{}  &   &   &   &                                                 &   \\
				\cite{20_dos_detection_prelimiary_work}                                           & O & X &   & \multicolumn{1}{p{2mm}|}{}  &   &   &   &   & \multicolumn{1}{p{2mm}|}{}  &   &   &   & \multicolumn{1}{p{2mm}|}{}  &   &   & \multicolumn{1}{p{2mm}|}{}  &   &   & X & X                                               &   \\
												
				\cite{28_intrusion_detection}                                                     & O & X & X & \multicolumn{1}{p{2mm}|}{}  &   &   &   &   & \multicolumn{1}{p{2mm}|}{}  &   &   & X & \multicolumn{1}{p{2mm}|}{}  &   &   & \multicolumn{1}{p{2mm}|}{}  & X & X &   &                                                 & X \\
				\cite{38_histogram_autoencoder}                                                   & O & X &   & \multicolumn{1}{p{2mm}|}{}  &   &   &   &   & \multicolumn{1}{p{2mm}|}{}  &   & X & X & \multicolumn{1}{p{2mm}|}{}  &   &   & \multicolumn{1}{p{2mm}|}{}  &   & X &   &                                                 & X \\
				\cite{65_fuse}                                                                    & O & X & X & \multicolumn{1}{p{2mm}|}{}  & X & X &   &   & \multicolumn{1}{p{2mm}|}{}  &   &   &   & \multicolumn{1}{p{2mm}|}{}  &   &   & \multicolumn{1}{p{2mm}|}{}  &   &   &   &                                                 &   \\
												
				\bottomrule 
			\end{tabular} }
	\end{center}
\end{table*}

\begin{table*}[!h]
				
	\caption{Anomaly detection and fault localization solutions (part 2)  }
    \label{table:anomalytablemulti-modal}
	\begin{center}
								
		\rowcolors{3}{gray!15}{white}
				
		\resizebox{\textwidth}{!}{
			\begin{tabular}{p{12mm} p{14mm}p{2mm}p{2mm}p{2mm}p{2mm}p{2mm}p{2mm}p{2mm}p{2mm}p{2mm}p{2mm}p{2mm}p{2mm}p{2mm}p{2mm}p{2mm}p{2mm}p{2mm}p{2mm}p{2mm}p{8mm}}
				\toprule
																
				\\

				\textit{Papers} & 
				\textit{Modality} & 
				\rotatebox{60}{\textit{Detection}} & 
				\rotatebox{60}{\textit{Localization}} & 
				\rotatebox{60}{\textit{Classification}} & 
				\rotatebox{60}{\textit{Memory (P1)}} &
				\rotatebox{60}{\textit{CPU (P2)}} & 
				\rotatebox{60}{\textit{Disk (P3)}} & 
				\rotatebox{60}{\textit{Deadlock (P4)}} &
				\rotatebox{60}{\textit{Process crash (P5)}}  &

				\rotatebox{60}{\textit{Message delay (A1)}} & 
				\rotatebox{60}{\textit{Configuration (A2)}} & 
				\rotatebox{60}{\textit{User interaction (A3)}}  & 
				\rotatebox{60}{\textit{IoT device (A4)}} &
																
				\rotatebox{60}{\textit{Message delay (C1)}} & 
				\rotatebox{60}{\textit{Configuration (C2)}} & 
				\rotatebox{60}{\textit{User interaction (C3)}}  &

				\rotatebox{60}{\textit{Insecure data (S1)}} &
				\rotatebox{60}{\textit{Service hijacking (S2)}} &
				\rotatebox{60}{\textit{External DoS (S3)}} &
				\rotatebox{60}{\textit{Internal DoS (S4)}} &
				\rotatebox{60}{\textit{Injection (S5)}} 
				\\   
				\midrule 

				\cite{2_LightGBM_fault_localizaiton}\cite{5_fault_localization_deep_learning}\cite{34_microdig} & M; T       & X & X & \multicolumn{1}{p{2mm}|}{X} & X & X & X &   & \multicolumn{1}{p{2mm}|}{X} & X                          &         &   &  \multicolumn{1}{p{2mm}|}{} &   &   &  \multicolumn{1}{p{2mm}|}{} &   &   &   &                     &         \\
				
				\cite{33_microrca}\cite{45_arvalus}                                                             & M; T       & X & X & \multicolumn{1}{p{2mm}|}{X} & X & X &   &   & \multicolumn{1}{p{2mm}|}{}  & X                          &         &   & \multicolumn{1}{p{2mm}|}{}  &   &   & \multicolumn{1}{p{2mm}|}{}  &   &   &   &                     &         \\
				\cite{19_fault_localization_semisupervised}                                                     & M; T       & X & X & \multicolumn{1}{p{2mm}|}{X} &   & X & X &   & \multicolumn{1}{p{2mm}|}{}  & X                          &         &   & \multicolumn{1}{p{2mm}|}{}  &   &   &  \multicolumn{1}{p{2mm}|}{} &   &   &   &                     &         \\
				\cite{26_multilayered_fault_detection}                                                          & M; T       & X & X & \multicolumn{1}{p{2mm}|}{X} & X & X & X &   & \multicolumn{1}{p{2mm}|}{} & X                          & X       &   & \multicolumn{1}{p{2mm}|}{}  &   &   & \multicolumn{1}{p{2mm}|}{}  &   &   &   &                     &         \\
				
				\cite{89_frl_mfpg}                                                                              & M; T       &   & X &  \multicolumn{1}{p{2mm}|}{} &   & X &   &   & \multicolumn{1}{p{2mm}|}{}  & X                          &         &   & \multicolumn{1}{p{2mm}|}{}  &   &   &  \multicolumn{1}{p{2mm}|}{} &   &   &   &                     &         \\
				\cite{83_microhecl}                                                                             & M; T       & X & X & \multicolumn{1}{p{2mm}|}{X} & X & X & X &   & \multicolumn{1}{p{2mm}|}{X} & X                          &         & X & \multicolumn{1}{p{2mm}|}{}  &   &   &  \multicolumn{1}{p{2mm}|}{} &   &   &   &                     &         \\
				\cite{86_tracegra}                                                                              & M; T       & X &   & \multicolumn{1}{p{2mm}|}{}  & X & X &   &   & \multicolumn{1}{p{2mm}|}{X} & X                          &         &   & \multicolumn{1}{p{2mm}|}{}  &   &   &  \multicolumn{1}{p{2mm}|}{} &   &   &   &                     &         \\
				\cite{78_tracerca}                                                                              & M; T       & X & X &  \multicolumn{1}{p{2mm}|}{} &   & X &   &   & \multicolumn{1}{p{2mm}|}{}  & X                          & X       & X & \multicolumn{1}{p{2mm}|}{}  &   &   &  \multicolumn{1}{p{2mm}|}{} &   &   &   &                     &         \\
				\cite{79_serviceanomaly}                                                                        & M; T       & X &   & \multicolumn{1}{p{2mm}|}{}  &   & X & X &   & \multicolumn{1}{p{2mm}|}{X} & X                          &         &   & \multicolumn{1}{p{2mm}|}{}  &   &   &  \multicolumn{1}{p{2mm}|}{} &   &   &   &                     &         \\
				\cite{81_dejavu}                                                                                & M; T       &   & X & \multicolumn{1}{p{2mm}|}{}  & X & X & X &   & \multicolumn{1}{p{2mm}|}{}  & X                          &         & X &  \multicolumn{1}{p{2mm}|}{} &   &   &  \multicolumn{1}{p{2mm}|}{} &   &   &   &                     &         \\

				\cite{72_trace_comparison}                                                                      & M; T       & X &   & \multicolumn{1}{p{2mm}|}{}  & X & X &   &   & \multicolumn{1}{p{2mm}|}{X} & X                          &         &   &  \multicolumn{1}{p{2mm}|}{} &   &   &  \multicolumn{1}{p{2mm}|}{} &   &   &   &                     &         \\

				\midrule

				\cite{23_fault_localization_log_metric}                                                         & M; L       & X & X & \multicolumn{1}{p{2mm}|}{}  & X & X & X & X & \multicolumn{1}{p{2mm}|}{X} & X                          & X       & X &  \multicolumn{1}{p{2mm}|}{} &   &   &  \multicolumn{1}{p{2mm}|}{} &   &   &   &                     &         \\
				\cite{35_deep_atentive_dam}                                                                     & M; L       & X &   & \multicolumn{1}{p{2mm}|}{X} & X & X & X &   & \multicolumn{1}{p{2mm}|}{X} & X                          &         &   &  \multicolumn{1}{p{2mm}|}{} &   &   & \multicolumn{1}{p{2mm}|}{} &   &   &   &                     &         \\
				\cite{52_mulan}                                                                                 & M; L       & X & X & \multicolumn{1}{p{2mm}|}{}  & X & X &   &   & \multicolumn{1}{p{2mm}|}{}  & X                          &         &   & \multicolumn{1}{p{2mm}|}{}  &   &   & \multicolumn{1}{p{2mm}|}{}  &   &   &   &                     &         \\
				\cite{68_kgroot}                                                                                & M; L       & X & X & \multicolumn{1}{p{2mm}|}{}  & X & X &   &   & \multicolumn{1}{p{2mm}|}{X} & X                          & X       &   & \multicolumn{1}{p{2mm}|}{}  &   &   &  \multicolumn{1}{p{2mm}|}{} &   &   &   &                     &         \\

				\midrule
				
				\cite{13_fault_localization_tracelog}\cite{15_fault_localization_unsupervised_trace}            & T; L       & X & X & \multicolumn{1}{p{2mm}|}{}  & X & X & X &   & \multicolumn{1}{p{2mm}|}{X} & X                          & X       & X &  \multicolumn{1}{p{2mm}|}{} &   &   &  \multicolumn{1}{p{2mm}|}{} & X &   &   &                     &         \\
				
				\cite{71_opentracing_detection}                                                                 & T; L       & X &   & \multicolumn{1}{p{2mm}|}{X} & X & X & X & X & \multicolumn{1}{p{2mm}|}{}  & X                          & X       &   &  \multicolumn{1}{p{2mm}|}{} &   &   &  \multicolumn{1}{p{2mm}|}{} &   &   &   &                     &         \\
				\cite{82_deeptralog}                                                                            & T; L       & X &   & \multicolumn{1}{p{2mm}|}{}  &   &   &   &   & \multicolumn{1}{p{2mm}|}{}  &                            & X       & X &  \multicolumn{1}{p{2mm}|}{} &   &   &  \multicolumn{1}{p{2mm}|}{} &   &   &   &                     &         \\
				\cite{73_yrca}                                                                                  & T; L       &   & X & \multicolumn{1}{p{2mm}|}{}  &   & X &   &   & \multicolumn{1}{p{2mm}|}{X} & X                          &         &   &  \multicolumn{1}{p{2mm}|}{} &   &   &  \multicolumn{1}{p{2mm}|}{} &   &   &   &                     &         \\
				
				\cite{49_latent_error}                                                                          & T; L       & X & X & \multicolumn{1}{p{2mm}|}{X} & X & X &   &   & \multicolumn{1}{p{2mm}|}{}  & X                          &         &   & \multicolumn{1}{p{2mm}|}{}  &   &   & \multicolumn{1}{p{2mm}|}{}  &   &   &   &                     &         \\
				
				\midrule
				
				\cite{57_instantOps}                                                                            & M; T; L    & X & X & \multicolumn{1}{p{2mm}|}{}  & X & X & X &   & \multicolumn{1}{p{2mm}|}{}  & X                          &         & X &  \multicolumn{1}{p{2mm}|}{} &   &   & \multicolumn{1}{p{2mm}|}{}  &   &   &   &                     &         \\
				\cite{14_fault_localization_GNN_multimodal}                                                     & M; T;  L   & X & X & \multicolumn{1}{p{2mm}|}{X} & X & X & X &   & \multicolumn{1}{p{2mm}|}{}  &                            &         &   & \multicolumn{1}{p{2mm}|}{}  &   &   & \multicolumn{1}{p{2mm}|}{}  &   &   &   &                     &         \\
				\cite{27_layered_vector_clock}                                                                  & M; T; L    & X & X & \multicolumn{1}{p{2mm}|}{X} & X & X & X &   &  \multicolumn{1}{p{2mm}|}{} & X                          &         &   &  \multicolumn{1}{p{2mm}|}{} &   &   &  \multicolumn{1}{p{2mm}|}{} &   &   &   &                     &         \\
				\cite{59_twin_graph}                                                                            & M; T; L    & X &   & \multicolumn{1}{p{2mm}|}{}  & X & X & X &   & \multicolumn{1}{p{2mm}|}{}  & X                          &         &   &  \multicolumn{1}{p{2mm}|}{} &   &   &  \multicolumn{1}{p{2mm}|}{} &   &   &   &                     &         \\
				\cite{61_eadro}                                                                                 & M; T; L    & X & X & \multicolumn{1}{p{2mm}|}{}  & X & X &   &   & \multicolumn{1}{p{2mm}|}{}  & X                          &         &   &  \multicolumn{1}{p{2mm}|}{} &   &   &  \multicolumn{1}{p{2mm}|}{} &   &   &   &                     &         \\
				\cite{80_anofusion}                                                                             & M; T; L    & X &   & \multicolumn{1}{p{2mm}|}{}  & X & X & X &   & \multicolumn{1}{p{2mm}|}{X} & X                          & X       & X &  \multicolumn{1}{p{2mm}|}{} &   &   &  \multicolumn{1}{p{2mm}|}{} &   &   &   &                     &         \\
				
				\midrule

				\cite{39_detective_dee}                                                                         & M; O       & X & X & \multicolumn{1}{p{2mm}|}{X} & X & X & X & X & \multicolumn{1}{p{2mm}|}{X} & X                          &         &   & \multicolumn{1}{p{2mm}|}{}  &   &   & \multicolumn{1}{p{2mm}|}{}  &   &   &   &                     &         \\
				\cite{85_TopoMAD}                                                                               & M; O       & X &   & \multicolumn{1}{p{2mm}|}{}  & X & X & X &   & \multicolumn{1}{p{2mm}|}{X} & X                          &         & X &  \multicolumn{1}{p{2mm}|}{} &   &   & \multicolumn{1}{p{2mm}|}{}  &   &   &   &                     &         \\
				
				\cite{58_murphy}                                                                                & M; O       &   & X & \multicolumn{1}{p{2mm}|}{X} & X & X & X &   & \multicolumn{1}{p{2mm}|}{} & X                          &         &   & \multicolumn{1}{p{2mm}|}{}  &   &   &  \multicolumn{1}{p{2mm}|}{} &   &   &   &                     &         \\
				
				
				\cite{62_SuanMing}                                                                              & M; T; O    & X & X & \multicolumn{1}{p{2mm}|}{}  & X & X & X &   &  \multicolumn{1}{p{2mm}|}{} & X                          &         &   &  \multicolumn{1}{p{2mm}|}{} &   &   & \multicolumn{1}{p{2mm}|}{}  &   &   &   &                     &         \\

				\cite{88_groot}                                                                                 & M; T; L; O &   & X &  \multicolumn{1}{p{2mm}|}{} &   & X & X &   & \multicolumn{1}{p{2mm}|}{X} & X                          &         & X &  \multicolumn{1}{p{2mm}|}{} &   &   & \multicolumn{1}{p{2mm}|}{}  &   &   &   &                     &         \\

				\bottomrule 
			\end{tabular} }
	\end{center}
\end{table*}

\subsection{Single-modal approaches}

Single-modal techniques primarily focus on a single type of data to determine the fault arising in the system. While some single-modal solutions still present competitive accuracy, authors of multi-modal solutions claim that only focusing on a single data type can reduce accuracy or limit the types of faults that may be detectable in the system (see \autoref{section: fault detection limitaitons}).

\subsubsection{Metrics}
The most often used data type for fault detection was metrics. This can be explained by the fact that the majority of the papers focus on detecting performance-related faults, which can most intuitively be diagnosed by examining component or system level measurements.

FSFP~\cite{9_fsps_fault_prediction} uses labeled runtime metrics to train a Long short-term memory neural network with the cross-attention mechanism. This model is then used to do a multi-label classification of faults within the system.

Yang and Jiang~\cite{10_indicator_prediciton} collect different kinds of data indicators and causal analysis using symbolic transfer entropy is performed to construct relationships between these indicators. Then convolutional and Long short-term memory neural networks are combined to forecast the values of these indicators. These forecasted values can then be used to predict faults and bottlenecks that the system might be facing, and prevent them before failures occur.

TADL~\cite{11_anomaly_detection_transformer} models both container relationships and temporal metrics relationships using a transformer. The transformer tries to reconstruct normal data, and if there is a reconstruction error, an anomaly is present. This anomaly is then localized using a container error scoring method.

Xie et al.~\cite{22_fault_localization_impact_graph_impacttracer} use timing metrics such as response time and success rate to detect if faults are occurring. Afterwards, they create an impact graph that analyses how much each node affects one another. Lastly, suspicion scores are calculated and the nodes are ranked based on how likely they are to have caused the fault.

FlowRCA~\cite{31_flowrca} uses a causality graph to trace the root cause of system faults. To detect anomalies, it uses the SPOT algorithm, which adapts to normal variations of metric fluctuations and distinguish anomalies. Then, these metrics are used to determine causal relationships between services to construct the causality graph. Finally, a random walk is applied to this graph to find services that occur the most in the path.

Kalinagac et al.~\cite{32_iot_root_cause} use Causal Bayesian Network to represent dependencies between microservices, and infer the faulty node within the system. A Causal Bayesian Network is trained on data that has been collected during fault injection to identify how services affect each other. To detect anomalies, they collect metric data and compare it to Service Level Agreement standards.

MicroCause~\cite{29_microcause_causality_inference} uses a path condition time series algorithm to capture the temporal data relationship. Afterward, a temporal cause-oriented random walk is performed, that uses both the temporal relationship graph and anomalous metrics that are detected. 

DyCause~\cite{36_dycause} uses a data collection proxy to collect kernel data of running services. This kernel data is then used by the SPOT algorithm to detect anomalies. Each local instance of the app creates a correlation graph, which is then fused into a single graph. Finally, the backward breadth-first search algorithm is applied from the front-end service to locate the faulty services by calculating their correlation to the faulty path and other services.

Zhang et al.~\cite{41_neural_multiclassification} use a neural transformation to transform metrics to increase data diversity. The original metric data is also converted into an adjacency matrix that holds Pearson correlation coefficient values between each data point. Finally, both of these data are fed into a Graph Neural Network that classifies the fault.

Xu et al.~\cite{42_tls_wgan_gp} use a generator to generate more class-specific samples. This avoids the problem of imbalanced data, where only a small subset of faults are labeled. Then, the discriminator is used to calculate the loss between generated and original data, and only accept data that resemble the original distribution. Finally, this data is fed into a classifier, which can be a support vector machine, random forest, etc. 

ASFC~\cite{64_asfc} is a module that is created to select the most appropriate model for fault localization. It trains multiple models on normal data and uses them to detect anomalies in future data. Using labeled data during the model selection phase, anomalies can be assigned labels to assist in diagnosis. Using the PC algorithm, the causal graph can be constructed and fault scores are calculated based on the node fault degree.

Jin et al.~\cite{43_component_analysis} use a sliding window to check if there are more than a third failing invocation chains. Once the threshold is passed, every node is traversed until no child nodes contain a failure. This is then combined with a single indicator anomaly detection, where key performance indicators are extracted, and trained on multiple classifiers to detect if there is a fault.

TopoRCA~\cite{47_topo_rca} first performs metric selection to more accurately locate the target anomalies. Then a decision tree is trained on this metric dataset to detect anomalies. Lastly, a topology graph is constructed, which represents the invocation relationship between service nodes and prunes the non-faulty nodes, and anomaly scores are calculated based on the node anomaly score and influence on other nodes.

LatentScope~\cite{50_latentscope} models root cause candidate nodes as latent variables, whose anomaly score is inferred from other metrics. This is then constructed into a dual-space graph that models observable and unobservable variables, modeling their relationships separately. Finally, localization is achieved by using a Regression-based Latent-space Intervention Recognition algorithm that uses linear regression to compute the error between the expected and observed latent variables.

Tritium~\cite{53_tritium} uses Service Level Objective violations to detect anomalies within services. Then Google's causal impact algorithm is used to infer the causal relationships and determine the root service.

SLA-VAE~\cite{56_semi_supervised_vae} extracts metric features and creates a probability density function to measure the divination of metrics. This data is then used to train a Variational Autoencoder, which is trained on labeled normal and anomalous data, but afterward performs active learning on small amounts of uncertain data identified by the ELBO ratio.

AutoMap~\cite{60_automap} samples metrics and creates an anomaly behavior graph. The graph is modeled by testing conditional independence on all pairs of services, where the weight of the edge is based on how large of an impact it has. Over time, the graph is aggregated to capture systems' behavior under normal conditions, and to detect anomalies, this normal behavior is subtracted from anomalies behavior to isolate it. Finally, past anomaly graphs are compared to current, updating the weights, and then performing heuristic random walk to pinpoint the root cause.

Nobre et al.~\cite{69_mlp} use Multi-layer perception to detect patterns in normal metric data. An anomaly can be detected if the provided date deviates from the historically learned pattern.

Ikram et al.~\cite{70_causal_discovery} use a modified PC algorithm to detect causal relationships between nodes. Once a change is detected in the modeled metrics, a Fault node is created in the causal graph that represents this anomaly. From this node neighborhood, the true fault case can be localized.

CausalRCA~\cite{84_causalrca} uses simple service objective violations to detect anomalies. Once they occur, it uses metrics to build a causal graph, which identifies anomaly propagation paths. To construct these graphs, gradient-based causal inference is used. Finally, PageRank is used to score the anomalous services and rank them.

MicroDiag~\cite{87_microdiag} collects metrics and uses the Distance-Based clustering algorithm BIRCH to detect Service Level Objective failures. Once an anomaly is found, using the collected component metrics a component dependency graph is constructed to model the anomaly propagation path. Using causal inference techniques, it generates a metrics causality graph that maps relationships between metrics. Finally, the system ranks metrics using a graph centrality algorithm to identify the most likely root cause of the anomaly.

ModelCoder~\cite{16_fault_localization_service_dependency_based} uses response times of requests to detect anomalies. Anomalous service data is collected and constructed into dependency graphs, from which feature vectors are extracted. These anomalous feature vectors are then compared to runtime feature vectors to calculate the likeliest service to have caused the fault.

TracerModel~\cite{17_fault_localization_continuation_of_16} is a follow-up paper that introduces TraceVAE and combines it with ModelCoder to improve anomaly detection, by using a variational autoencoder, which is trained on normal data to calculate the expected latency of services. This expected latency is then used to determine thresholds to detect faulty services.

\subsubsection{Logs}
Logging message formats are not always consistent among system components, making these methods more difficult to implement in preexisting architectures. This may explain the lack of authors exploring single-modal logging methods for fault detection.

LogRAG~\cite{37_lograg} uses logs to detect anomalies. First logs are collected and parsed to have the appropriate representation. Afterward, they are fed to a deep neural network to learn to identify normal data. To detect anomalies, this network is used as a primary classifier, and to decrease the number of false positives, a large language model such as ChatGPT is used to reevaluate the anomalous logs.

Aggarwal et al.~\cite{40_causal_slo} monitor the system for log errors. Once a certain threshold is reached, an anomaly is assumed, and appropriate error logs are collected to perform fault localization. The logs are then modeled in time series which are used to construct a causal dependency graph using the PC algorithm. Finally, the anomaly score is calculated using the PageRank algorithm, which traverses the causal graph using a random walk.

Aggarwal et al.~\cite{67_golden_signals} use log errors to detect anomalies within the system. Once a certain error threshold is reached, logs are modeled as multivariate time series data, from which a causal relationship graph can be inferred using Granger causality techniques. Finally, PageRank is used to score and localize errors.

LogAttention~\cite{74_logattention} parses logs to extract important information. These parsed logs are then fed into 3 different classifiers: heuristic, supervised, and unsupervised classifier. All their outputs are then supplied to an ensemble model, which appropriately weighs the classifiers based on their accuracy and outputs an anomaly score indicating the severity of the fault.

\subsubsection{Traces}

Traces are a good way to identify component dependency, but usually do not contain any information about faults within its structure, only relying on abnormalities in the traffic to detect faults. However, this modality does get more utilized in methods using multiple data types.

Bogatinovski~\cite{46_self_supervised} first performs Masked Span Prediction, which masks a random event within a trace and tries to predict it using the available information. This model is then used to generate expected traces, and compares them to real traces to identify anomalies.

Informer~\cite{76_informer} uses Remote Procedure Calls (RPC) to build an RPC graph that models the dependencies of services. These graphs are grouped into clusters using density-based clustering to identify RPC chain patterns of highly related RPCs. Lastly, Diffusion Convolution Recurrent Neural Network is trained on RPC graphs to predict RPC traffic and detect anomalies.

Wang et al.~\cite{77_workflow_aware} characterize traces into call trees, which are then grouped into similarity clusters using the tree edit distance algorithm. These clusters are then used as a baseline to detect workflow anomalies. To detect performance anomalies, the coefficient of variance is calculated to measure timing variability in services, and Principal Component Analysis is performed to localize the faulty services.

MicroNet~\cite{90_micronet} uses traces to construct a microservice-operation invocation network. This graph is then used to generate meta calls, which are evaluated for abnormal latency to detect anomalies. Finally, anomalies are backtracked and root cause candidates are ranked using a modified PageRank algorithm.

GTrace~\cite{54_groupwise} splits trace data into groups to better isolate trace characteristics. This data is then fed into a Variational Autoencoder model, which attempts to model the normal data. If the test sample deviates too much from the predicted data, the anomaly is detected.

\subsubsection{Miscellaneous}

Techniques that utilize other data types for fault localization tend to be more focused on detecting specific kinds of system faults, and as can be seen in \autoref{table:anomalytablesinglemodal}, predominantly targeting security related faults. Additionally, some of these methods are made specifically for IoT environments to detect IoT failure (\textbf{A4}).

H-FaTMA~\cite{1_IoT_hybrid_FT} is a comprehensive framework that provides both fault detection and recovery mechanisms. Things Reactive Monitor examines IoT devices for any anomalies in the data by comparing them with other devices or standardized values, while Things Proactive Monitor using real-time data predicts upcoming faults. To keep the Things Proactive Monitor accurate, it is trained in the cloud with historical data. Fault recovery of this framework is discussed in \autoref{section: fault recovery}.

Alvarez et al.~\cite{6_fault_tolerance_telerehab_iot} propose a simple framework for healthcare machine fault detection. It uses fuzzy logic to detect anomalies within the IoT devices and alerts the user or shuts down the system in critical scenarios.

Flora et al.~\cite{28_intrusion_detection} collect low-level system traces of running microservice containers under normal circumstances. This data is then used to create a profile which is then compared to the runtime profile using algorithms and classifiers to detect anomalies during runtime.

Kotenko et al.~\cite{38_histogram_autoencoder} represent system-level calls as histograms, which are then fed into an autoencoder. First, the data under normal operation is fed to train the autoencoder. This enables the autoencoder to reconstruct the data, and if it differs from the current data, an anomaly has occurred.

FUSE~\cite{65_fuse} uses eBPF technology to see microservice runtime behavior on a kernel level. This allows for the creation of a unique hash-based digest for each microservice invocation. The digest can then be compared to a historical digest to see if a fault has occurred.

Castro et al.~\cite{20_dos_detection_prelimiary_work} use attribute aggregation to compute a global score that indicates if the system is under attack or not. While this is a preliminary paper, they are able to achieve precision and recall higher than 85\% on their own dataset, which could be further improved.

\subsection{Multi-modal approaches}

Multi-modal methods try to utilize different types of data by fusing them to create more informative feature vectors, or using each data type for a particular analysis of fault causes and propagation.

\subsubsection{Metrics and traces}
Combining metrics and traces seems to be the most frequently adopted method. Metrics are typically used for detecting component degradation or anomalous behavior, while traces help to construct correlation graphs amongst service nodes.

LightGBM~\cite{2_LightGBM_fault_localizaiton} uses a lightweight version of the gradient lifting tree with four new optimizations that increase the model's training speed. The model is trained on different metrics and trace data and then used to locate and identify faults in the system.

Chen et al.~\cite{5_fault_localization_deep_learning} uses a request-weighted graph to characterize the actions taken between microservices. These traces are collected under normal and abnormal circumstances, and discrepancies within the request-weighted graph are identified as anomalies. Afterward, the request graph is used to train a deep neural network model to find the root cause of the observed anomaly.

Li et al.~\cite{19_fault_localization_semisupervised} uses a semisupervised method for fault detection. It splits the data into time series to account for fluctuation in time and uses previously collected labeled and unlabeled data to initialize the cluster centers for classifying both normal and abnormal data.

Wang et al.~\cite{26_multilayered_fault_detection} uses a multilayered approach, where fault detection happens on service, resource and metric layers. With this approach, the anomaly can be periodically localized, first by finding the affected service, then the related containers, and then the underlying resource failure.

MicroDig~\cite{34_microdig} uses Service Level Objective standards to detect anomalies occurring within the application. Associated calls that caused the anomaly are identified, from which the heterogeneous propagation graph is constructed, which describes the causal relationships between both upstream and downstream nodes. Then, a modified random walk is used to identify the faulty node.

MicroRCA~\cite{33_microrca} collects live metrics from the application and system level and trains an unsupervised clustering algorithm to identify anomalies within the data. To localize faults, first attribute graph is constructed which models anomaly propagation along the service call path, then a subgraph is extracted that contains the nodes experiencing anomalies, and finally, the node anomaly score is calculated and faulty nodes ranked.

Arvalus~\cite{45_arvalus} uses key performance metrics to create a dependency graph between microservices. It then uses a Graph Convolutional Neural Network to feed in the data to perform fault localization.

BARO~\cite{55_baro} uses the Multivariate Bayesian Online Change Point Detection method to model the dependency of metric data expressed as time series. Once the algorithm detects a significant change in the time series, nonparametric hypothesis testing is performed to identify candidates and rank them based on how much the metric deviates from the norm.

Meng et al.~\cite{72_trace_comparison} collects runtime traces and characterizes them using relevant metrics. Using the collected baseline of normal data, faults are classified into either trace structural anomalies by comparison tree edit difference algorithm or response time anomalies by utilizing principal component analysis.

TraceRCA~\cite{78_tracerca} uses an unsupervised multi-metrics anomaly detection model that first extracts the useful features related to the currently occurring anomaly and then detects invocations based on these features. Afterward, suspicious microservice sets are found by analyzing how many times they contribute to anomalous traces. Finally, anomalous services are ranked based on their individual contribution to the anomaly within their service set.

ServiceAnomaly~\cite{79_serviceanomaly} creates a Context Propagation Graph, which models user request propagation among the services. Using additional metrics to characterize the graph, anomalies can be detected by comparing their traces to the Context Propagation Graph.

DejaVu~\cite{81_dejavu} uses historical fault incidents to construct a failure dependency graph that maps dependencies between all faulty node candidates. For each candidate, their associated metric data is then transformed into a feature vector, which is then integrated with the failure dependency graph using Graph Attention Network. This final aggregated feature vector is then fed to a dense neural network, to assign an anomaly score to each faulty node candidate.

MicroHECL~\cite{83_microhecl} uses traces and metrics to construct a service call graph, once an anomaly is detected by either performance, reliability, or traffic anomaly detection models. From the initially reported faulty node, backward traversal is performed to find any candidate nodes that may have propagated the anomaly. Finally, the Pearson correlation coefficient is used to measure the similarity between the initial candidate and all others and rank them based on the anomaly score.

TraceGra~\cite{86_tracegra} uses metrics and traces to construct a Trace Propagation Graph. By encoding traces into vectors, they can be clustered based on their Euclidean distance similarity. Redundant features are removed, and a Variational Graph Autoencoder along with an LSTM Autoencoder are used to extract spatial and temporal features. Finally, observed data is classified as anomalous if its reconstruction error is too high.

FRL-MFPG~\cite{89_frl_mfpg} uses microservice Key Performance Indicators and metrics to detect anomalies. Afterward, the link call graph is constructed according to the abnormal request call data. Then, using historical fault data and the corresponding events, a microservice fault correlation directed graph is constructed. Finally, both of these graphs are combined, and a modified random walk is used to traverse the graph and localize the root cause.

\subsubsection{Metrics and logs}

Techniques using metrics and logs tend to either fuse both data types together or use logs to construct correlation graphs, similar to how it is performed using traces.

Zhang et al.~\cite{23_fault_localization_log_metric}  use the PC algorithm to detect the causal relationships of services with system logs and builds a service dependency graph with the causal relationships. Then, an anomaly detection method with system logs and monitoring metrics is used. By transforming system logs into frequent signals, they use deep learning (transformer) to capture the variant features of system logs and monitoring metrics. For each service in the anomalous service set, a depth-first traversal is performed on the dependency graph starting from the anomalous service, and all the traversed paths are collected. Then by calculating how many times a node appears on these paths, the nodes are ranked based on their likelihood to produce the fault.

DAM~\cite{35_deep_atentive_dam} uses logs and metrics are fused to construct input data for the Long short-term memory model. This model is then used to predict the metrics of the system, and if divination from the norm is detected, the anomaly can be identified.

MULAN~\cite{52_mulan} trains and uses a log tailored language model to extract meaningful information from log data. Then, modality specific and invariant data is extracted both from parsed logs and metric data. Finally, modality data is fused together into a causality graph, which is traversed with a random walk to locate the root cause.

KGroot~\cite{68_kgroot} uses logs and metrics and transforms them into structured events. These events are then used to discover causal relationships, which are used to construct knowledge graphs. These knowledge graphs are built using historic fault data, and then compared to online graphs using Graph Convolutional Network to detect and localize faults.

\subsubsection{Traces and logs}

Log and trace combination has a strong resemblance to metric and trace combination, where instead of using metrics to describe the service node's behavior, logs are used instead.

Sun et al.~\cite{13_fault_localization_tracelog} collect traces and logs of both normal and abnormal services. Suspicion scores are calculated by analyzing the similarity of obtaining trace logs using the Levenshtein distance or cosine similarity. Finally, the possible culprits are ranked.

FSF~\cite{15_fault_localization_unsupervised_trace} monitors HTTP errors within the traces between microservices. If an error is detected, causal inference is applied to determine the services that caused the fault. Lastly, logs during the error period are collected from the suspected services to perform manual analysis.

Khanahmadi et al.~\cite{71_opentracing_detection} uses OpenTracing to collect traces to construct a service dependency graph and extract high-level information from logs. This is then used to detect performance and service dependency anomalies. To detect fault types, multiple machine learning algorithms can be trained on labeled data. Finally, faults can be localized by mapping anomalies to span trace data.

yRCA~\cite{73_yrca} uses logs and collected events to perform root cause analysis. This method however requires a specific logging structure to be employed, such that the localization can take place.

DeepTraLog~\cite{82_deeptralog} parses traces and logs, extracting span relationships from traces and event information from logs. These are then turned into vectors and Trace Event Graphs and fed into a Gated Graph Network model, to train it to predict latent Trace Event Graph representation. Using Support Vector Data Description, anomalous traces can be detected.

MEPFL~\cite{49_latent_error} uses system trace logs and transforms them into feature vectors. These vectors are collected during normal runtime and fault-injected runtime. These feature vectors are then used to train multiple models that each perform a specific type of classification: trace error, faulty microservice and fault type predictions.

\subsubsection{Metrics, logs and traces}

Methods that use metrics, traces, and logs build on top of previously mentioned techniques. It either fuses all data types to create a feature vector or uses traces to construct correlation graphs and assign attributes to these nodes based on log and metric data.

DiagFusion~\cite{14_fault_localization_GNN_multimodal} builds a dependency graph using traces and deployment data. Then using logs, metrics and traces, a feature vector is constructed that combines all of these data types using a fastText machine learning model. Finally, a Graph Neural Network (GNN) combines the graph and vector to perform fault localization and classification.

Bento et al.~\cite{27_layered_vector_clock} uses a multilayered approach, where the first layers independently try to detect anomalies that either break service level constraints or application level constraints. After the anomaly is detected, causality is inferred using vector clocks, and the most likely culprits are ranked by how close they are to the observed anomaly.

InstantOps~\cite{57_instantOps} fuses logs, metrics, and traces into a time-series format, which then is stitched together and used to construct a dependency graph. Finally, this graph is then fed to a Graph Neural Network to localize faults, and Gated Recurrent Unit model is utilized to capture systems changing dynamics over time.

MSTGAD~\cite{59_twin_graph} uses logs, metrics, and traces to construct a relationship graph amongst the microservice nodes. This graph is then used to train a transformer-based neural network, that models the normal data and is able to detect anomalies.

Eadro~\cite{61_eadro} uses logs, metrics, and traces to learn intra-service behaviors using a different approach for each type of data. To construct a dependency graph, traces are used to identify relationships, and log and metric data are embedded into the nodes. Then, Graph Neural Network is used to capture the dependency aware features, which are then used to identify anomalies and localize them.

AnoFusion~\cite{80_anofusion} takes logs, metrics, and traces and turns them into time series. Using these data streams, a heterogeneous graph is built, which is then fed into a graph transformer network and Graph Attention Network to find relationships between the time series and extract meaningful features. Finally, these features are fed into a Gated Recurrent Unit, to predict the expected system state, and compare it to the live system state to detect anomalies.

\subsubsection{Miscellaneous}

Miscellaneous data types can be used to increase accuracy or detect certain types of faults. The observed additional data types are user request metadata, systems topology, or syscalls made within the service node.

Detective-Dee~\cite{39_detective_dee} monitors system metrics and syscalls to detect any anomaly in the system. It uses a variation of the Compressed Sensing algorithm with additional computational optimizations to reconstruct the metric signals and compares if they deviate from the measurements more than the white noise. To localize the faults, it uses instrumentation and static analysis to identify vulnerable functions within the code that may have caused the fault in the system.

Murphy~\cite{58_murphy} is a performance diagnosis system, that uses known metrics to localize the root cause of performance faults in the system. It uses metric metadata and topology to describe service relationships, which then are used to construct a Markov Random Field. Markov Random Fields are constructed live using the previous week's data since the relationships between metrics can change over the system runtime period. Lastly, Gibbs sampling is used to infer the root cause by determining how significantly metrics contribute to a cause.

TopoMAD~\cite{85_TopoMAD} collects metric and topology data. This data is then used to train a GraphLSTM model, which is constructed using Graph Neural Network and Graph Attention Network. Finally, this model is used to reconstruct the observed data and compare it to detect anomalies.

SuanMing~\cite{62_SuanMing} is able to predict performance degradation before it even happens. It uses multiple models to estimate user requests and systems performance based on metrics, traces, and user demand. Cascading all these models, SuanMing can predict service level performance.

Groot~\cite{88_groot} uses traces and logs to maintain a dependency graph. Afterward, this graph is combined with metrics and developer activities to generate an event causality graph. Finally, using a customized PageRank algorithm, anomaly scores are calculated and ranked.

\subsection{Limitations}
\label{section: fault detection limitaitons}

Only a small amount of papers engaged in fault-oriented performance analysis of their systems~\cite{77_workflow_aware, 84_causalrca, 87_microdiag, 89_frl_mfpg}. A lot of papers did not explicitly mention what kind of faults they were injecting~\cite{11_anomaly_detection_transformer, 22_fault_localization_impact_graph_impacttracer, 36_dycause, 37_lograg, 42_tls_wgan_gp, 45_arvalus, 46_self_supervised, 49_latent_error, 50_latentscope}, or briefly mentioned them but did not analyze the influence of fault class on the methods detection accuracy~\cite{78_tracerca, 79_serviceanomaly, 80_anofusion, 82_deeptralog, 83_microhecl, 85_TopoMAD, 86_tracegra, 88_groot, 90_micronet, 81_dejavu}.

Very few papers discuss fault localization when multiple faults happen concurrently. As systems grow larger, simultaneous faults may become more prevalent, so this topic should be more explored. Since a lot of papers perform ranking of candidate root cause nodes~\cite{13_fault_localization_tracelog, 22_fault_localization_impact_graph_impacttracer, 23_fault_localization_log_metric, 33_microrca, 40_causal_slo, 55_baro}, multiple faulty nodes may show higher in the list, but there is no distinction between false positives and multiple faulty nodes.

Papers that incorporate multi-modal data in their method claim that using multi-modal data is important to catch as many faults as possible, since each fault can exhibit different kind of features that could be exclusive to either logs, metrics, or traces~\cite{23_fault_localization_log_metric, 52_mulan,61_eadro}, or may be easier to detect by combining multiple data types~\cite{57_instantOps, 59_twin_graph}.

The majority of papers used their own custom datasets for evaluating their method. While it helps to showcase their accuracy in discovering faults, it makes it difficult to compare them with other solutions. Some papers used standardized evaluation datasets, such as AIOps Challenge 2020~\cite{16_fault_localization_service_dependency_based,17_fault_localization_continuation_of_16, 19_fault_localization_semisupervised, 26_multilayered_fault_detection, 43_component_analysis} or Sock-Shop Dataset~\cite{11_anomaly_detection_transformer, 26_multilayered_fault_detection, 31_flowrca}.

Many papers rely on supervised machine learning models to detect or localize faults~\cite{5_fault_localization_deep_learning, 9_fsps_fault_prediction, 11_anomaly_detection_transformer, 12_automated_monitoring_lstm, 14_fault_localization_GNN_multimodal,17_fault_localization_continuation_of_16, 26_multilayered_fault_detection, 31_flowrca, 32_iot_root_cause, 41_neural_multiclassification, 43_component_analysis, 46_self_supervised, 47_topo_rca, 49_latent_error}. While efficiently detecting labeled faults, it may cause trouble for faults that are not in the training set, or are incorrectly labelled~\cite{15_fault_localization_unsupervised_trace}.

These methods have different levels of granularity in identifying the type of faults. The majority of papers do not perform node classifications and localize the faults to either services or containers. However, some papers include fault classification in their localization models~\cite{5_fault_localization_deep_learning,9_fsps_fault_prediction, 17_fault_localization_continuation_of_16, 19_fault_localization_semisupervised}, and in some cases having comparable accuracies to localization~\cite{5_fault_localization_deep_learning, 19_fault_localization_semisupervised}; 

An overwhelming amount of detection methods are centered around detecting faults in any microservice deployment scheme, as long as it supports the necessary data types needed for fault detection. A very small number are specifically targeted towards IoT environments~\cite{1_IoT_hybrid_FT, 32_iot_root_cause, 6_fault_tolerance_telerehab_iot}.

\section{Fault recovery}
\label{section: fault recovery}

Fault recovery is a complex task that is most often performed manually. The presence of false positives and negatives in fault detection algorithms makes recovery heavily rely on their accuracy for meaningful results. While the literature is more focused on detection, there still exist some methods that attempt to automatically recover from certain kinds of faults the system may experience.
An overview of analyzed papers, the type of solution they propose, and fault coverage are summarized in \autoref{table:recoverytable}. The next sections give a brief explanation of the methods used in these papers.

\begin{table*}[!h]
					
	\caption{Recovery solutions  }
	\label{table:recoverytable}	
	\begin{center}
										
		\rowcolors{3}{gray!15}{white}
						
		\resizebox{\textwidth}{!}{
			\begin{tabular}{p{4mm} c p{2mm}p{2mm}p{2mm}p{2mm}p{2mm}p{2mm}p{2mm}p{2mm}p{2mm}p{2mm}p{2mm}p{2mm}p{2mm}p{2mm}p{2mm}p{2mm}p{8mm}}
				\toprule
																				
				\\

				\textit{Papers} & 
				\textit{Solution method} & 
				
				\rotatebox{60}{\textit{Memory (P1)}} &
				\rotatebox{60}{\textit{CPU (P2)}} & 
				\rotatebox{60}{\textit{Disk (P3)}} & 
				\rotatebox{60}{\textit{Deadlock (P4)}} &
				\rotatebox{60}{\textit{Process crash (P5)}}  &

				\rotatebox{60}{\textit{Message delay (A1)}} & 
				\rotatebox{60}{\textit{Configuration (A2)}} & 
				\rotatebox{60}{\textit{User interaction (A3)}}  & 
				\rotatebox{60}{\textit{IoT device (A4)}} &
																				
				\rotatebox{60}{\textit{Message delay (C1)}} & 
				\rotatebox{60}{\textit{Configuration (C2)}} & 
				\rotatebox{60}{\textit{User interaction (C3)}}  &

				\rotatebox{60}{\textit{Insecure data (S1)}} &
				\rotatebox{60}{\textit{Service hijacking (S2)}} &
				\rotatebox{60}{\textit{External DoS (S3)}} &
				\rotatebox{60}{\textit{Internal DoS (S4)}} &
				\rotatebox{60}{\textit{Injection (S5)}} 
				\\   
				\midrule 

				\cite{12_automated_monitoring_lstm}     & \multicolumn{1}{c|}{reconfiguration; circuit breaking} & X & X & X &   & \multicolumn{1}{c|}{}  &   &   &   &  \multicolumn{1}{c|}{}                      &                                                     &   & \multicolumn{1}{c|}{}  &   &   &   &   &   \\
				\cite{18_versioning_fault_localization} & \multicolumn{1}{c|}{reconfiguration}                   & X & X & X &   &  \multicolumn{1}{c|}{} & X & X &   &  \multicolumn{1}{c|}{}                      &                                                     &   & \multicolumn{1}{c|}{}  &   &   &   &   &   \\
				\cite{21_automatic_recovery_microras}   & \multicolumn{1}{c|}{reconfiguration}                   & X & X &   &   & \multicolumn{1}{c|}{}  &   &   &   &  \multicolumn{1}{c|}{}                      &                                                     &   & \multicolumn{1}{c|}{}  &   &   &   &   &   \\
				\cite{75_misuse_detection}              & \multicolumn{1}{c|}{reconfiguration}                   &   &   &   &   & \multicolumn{1}{c|}{}  &   &   &   &  \multicolumn{1}{c|}{}                      &                                                     &   & \multicolumn{1}{c|}{}  & X & X &   &   & X \\
				\cite{30_redundancy_cloud,63_kubernetes}              & \multicolumn{1}{c|}{replication}                       &   &   &   &   & \multicolumn{1}{c|}{X} &   &   &   &  \multicolumn{1}{c|}{}                      &                                                     &   & \multicolumn{1}{c|}{}  &   &   &   &   &   \\
				\cite{1_IoT_hybrid_FT}                  & \multicolumn{1}{c|}{replication;  circuit breaking}    &   &   &   &   &  \multicolumn{1}{c|}{} &   &   &   & \multicolumn{1}{c|}{X}                      &                                                     &   & \multicolumn{1}{c|}{}  &   &   &   &   &   \\
				\cite{51_circuit_breaking,24_security_traffic_routing}              & \multicolumn{1}{c|}{circuit breaking}                  &   &   &   &   & \multicolumn{1}{c|}{}  &   &   &   &  \multicolumn{1}{c|}{}                      &                                                     &   & \multicolumn{1}{c|}{}  &   &   & X & X &   \\
				\cite{66_circuit_breaker}               & \multicolumn{1}{c|}{circuit breaking}                  &   &   &   &   & \multicolumn{1}{c|}{}  &   &   & X &  \multicolumn{1}{c|}{}                      &                                                     &   & \multicolumn{1}{c|}{}  &   &   & X & X &   \\
				\cite{44_smaac}                         & \multicolumn{1}{c|}{dynamic access control}            &   &   &   &   & \multicolumn{1}{c|}{}  &   &   & X & \multicolumn{1}{c|}{}                       &                                                     &   & \multicolumn{1}{c|}{}  & X & X &   &   & X \\
				\cite{48_runtime_trust}                 & \multicolumn{1}{c|}{dynamic access control}            &   &   &   &   & \multicolumn{1}{c|}{}  &   &   & X & \multicolumn{1}{c|}{}                       &                                                     &   & \multicolumn{1}{c|}{}  & X & X &   &   &   \\
				\cite{4_fog_framework}                  & \multicolumn{1}{c|}{framework}                         & X & X & X &   & \multicolumn{1}{c|}{X} & X &   &   & \multicolumn{1}{c|}{}                       &                                                     &   & \multicolumn{1}{c|}{}  &   &   &   &   &   \\
				\cite{25_micro_intrusion_recovery}      & \multicolumn{1}{c|}{action reversion}                  &   &   &   &   & \multicolumn{1}{c|}{}  &   &   & X & \multicolumn{1}{c|}{}                       &                                                     &   & \multicolumn{1}{c|}{}  & X & X &   &   & X \\

				\bottomrule 
			\end{tabular} }
	\end{center}
\end{table*}

\subsection{Reconfiguration}

Reconfiguration methods attempt to put a system in a consistent state by configuring certain system or component parameters.  
Hu et al.~\cite{12_automated_monitoring_lstm} incorporate circuit breaking and real-time configuration of services. The decision made uses the knowledge graph that can be constructed by training Long short-term memory neural networks on historical data.
VMAMV~\cite{18_versioning_fault_localization} creates a service dependency graph that is able to detect errors related to changing API specifications. In cases where a specific version of a node is required, VMAMVS is able to add or remove nodes from the system.
Wu et al.\cite{21_automatic_recovery_microras} model the system using its attributes, and try to estimate how an action may impact the system state, by evaluating its benefit and associated risk. Using fuzzy logic, the impact of each action is calculated, and the most positive action is performed.
Aly Amin et al.~\cite{75_misuse_detection} use the Dynamic Topology Adjustment operator inside Kubernetes to modify the network topology, to avoid malicious attacks.

\subsection{Replication}

Replication of nodes is a common technique used in microservice architecture to ensure resistance against service crashes. While replication is a straightforward solution, determining the optimal number of replicas can be challenging. 
O’Neill and Soh~\cite{30_redundancy_cloud} designed a utility function that estimates the appropriate amount of redundant nodes necessary to ensure fault tolerance and minimize hosting costs. It uses the probability of container failure to estimate the expected value of monetary loss, and compares it to the marginal cost of adding extra nodes to find the best amount of active redundant services.
Vayghan et al.~\cite{63_kubernetes} introduce a state controller, which does passive replication of nodes, where there is an active node performing work, and a passive node that performs state replication and is ready to take over in case of the active nodes failure.

\subsection{Circuit breaking}

Circuit breakers are components that sit between service nodes, and drop incoming or outgoing requests in an attempt to avoid propagating errors, or reduce the load for overburdened service nodes.
Sedghpour et al.~\cite{51_circuit_breaking} use adaptive circuit breaking that measures the 95th percentile of response time and queue size. Using an exponential smoothing function on these metrics gives an indicator of when to activate the circuit breaker.
H-FaTMA~\cite{1_IoT_hybrid_FT}, after detecting a fault, uses a biographical model to formalize reaction rules to the system whenever a fault is detected. Most of the faults are handled by circuit breaking and replication.
Alboqmi et al.~\cite{24_security_traffic_routing} use information about incoming requests to decide how to route the request to not overwhelm the system. In cases where a certain threat threshold is reached, the request is blocked.   
Hlybovets and Paprotskyi~\cite{66_circuit_breaker} adapt the circuit-breaking pattern to use more dynamic methods to control the state of the circuit. Using system metrics, the needed time for opening the circuit can be more reliably estimated, and system downtime can be avoided.

\subsection{Dynamic access control}

Dynamic access control manages access for specific users or service nodes to avoid malicious attacks or putting the system in an illegal state. These solutions tend to target more security related faults.
SMAAC~\cite{44_smaac} uses shared threat intelligence data to verify if the actions taken by a service are allowed. If not, a dynamic access control policy is implemented to restrict illegal actions.
Alboqmi et al.~\cite{48_runtime_trust} introduce a runtime trust evaluator that evaluates the trustworthiness of other service nodes, based on derived service dependency graph and service connectivity. It assigns each action a cost, and once the threshold is reached, the requesting service is denied access.

\subsection{Miscellaneous}

Framework and action revision techniques were additionally identified but do not fit into any larger category. 
Whaiduzzaman et al.~\cite{4_fog_framework} creates a fault tolerance framework for fog computing environments. Its main contribution is introducing redundancy in the master fog node infrastructure, and coordinating between the backup master fog nodes to synchronize the recovery of the system.
$\mu$Verum\cite{25_micro_intrusion_recovery} allows for intrusion recovery, where a user has performed illegal actions within the system. To achieve this, $\mu$Verum models the dependency graph and traces back each affected microservice, reversing all malicious actions that were performed.

\subsection{Limitations}

As previously mentioned, these methods still heavily rely on the accuracy of fault localization algorithms~\cite{21_automatic_recovery_microras}. Because of this, some methods still ask for user confirmation before performing changes to the system~\cite{25_micro_intrusion_recovery}.
Since services get updates independently, recovery mechanisms may not be applicable after the integrated changes~\cite{21_automatic_recovery_microras}. This means that developers have to take into account the recovery procedure when updating their services, or running multiple versions simultaneously.
When it comes to recovering from performance related faults, recovery mechanisms are only slightly better than just restarting the affected node~\cite{21_automatic_recovery_microras}.

\section{Fault coverage}
\label{section: fault coverage}

This section discusses the faults that current detection and recovery mechanisms cover. Additionally, notable papers that show high accuracy or reliability are discussed.

\subsection{Performance related faults}

The majority of detection methods target performance related faults. Having high availability is usually one of the goals that microservice architecture tries to achieve, which could explain why so much research has been done in this area. Additionally, a significant number of papers still struggle to achieve accuracies above 90\% on their own test sets, meaning that there is still room for improvement. The only outlier was deadlock, which could be explained by the fact that it could be bundled with user interaction faults and not explicitly separated.
For detecting performance related faults, TADL~\cite{11_anomaly_detection_transformer} has shown high accuracies on the Sock-Shop dataset, and showed improvement when compared to other methods, such as TopoMAD~\cite{85_TopoMAD}. FSFP~\cite{9_fsps_fault_prediction} has also shown high accuracy, and has been tested on a larger range of faults.
When it comes to recovery algorithms, it also lacks mechanisms for recovering from deadlocks. The solutions for system faults usually involve replication, circuit breaking or reconfiguration. Circuit breaking tends to be incorporated together with reconfiguration and replication mechanisms. This suggests that recovery mechanisms may use multiple techniques to ensure fault tolerance.
A general recovery solution is presented by Wu et al~\cite{21_automatic_recovery_microras} where different recovery approaches are suggested, and by performing performance estimations, the configuration with the highest benefit is chosen.

\subsection{Architecture related faults}

Detection algorithms overwhelmingly focus on detecting message delays in comparison to other faults in this group. This can be again explained by the focus on the availability of the microservice architecture, and that this has been documented to be a common fault occurring in this architecture. IoT device failure does not have much coverage in this survey mostly because of the exclusion of IoT architecture from the search query, since it does not always directly relate to microservice architecture.
Notable papers for fault detection are those of Wang et al.~\cite{26_multilayered_fault_detection} for its high accuracy and inclusion of fault classification, and AnoFusion~\cite{80_anofusion} for its high accuracy in anomaly detection.
Recovery methods tend to again have multiple recovery techniques for handling architecture related faults. Dynamic access control has shown to be useful for mitigating user interaction faults by dynamically changing their action privileges. For reducing message delay or the impact of configurations on the system, reconfiguration seems to be the most straightforward solution. 
Notable papers are VMAMSVS~\cite{18_versioning_fault_localization} which monitors multiple versions of microservices running, and can adjust the active versions based on runtime analysis. Both SMAAC~\cite{44_smaac} and Alboqmi et al.~\cite{48_runtime_trust} use dynamic access control to adjust user actions, to reduce the harm done to the system.

\subsection{Component related faults}

From the surveyed literature, none of the detection or recovery mechanisms cover component related faults, such as faults present in service registry or monitoring mechanism. These are important components for ensuring communication and reliability in microservice architecture, so their malfunction can cause the entire system to fail. 

\subsection{Security related faults}

Several frameworks combine both detection and recovery inside a single framework, such as $\mu$Verum~\cite{25_micro_intrusion_recovery}, SMAAC~\cite{44_smaac} and R. Alboqmi et al.~\cite{48_runtime_trust}. This is due to them using specific data for decision making that most detection methods do not focus on.
Flora et al.~\cite{28_intrusion_detection} collect low-level system traces, and using them can detect if a service node has an intrusion. For detecting denial of service attacks, a promising approach is proposed by Castro et al.~\cite{20_dos_detection_prelimiary_work} who have shown that their method can achieve high accuracy in its preliminary stage.
For regulating traffic, circuit breakers seem to be the dominant method of recovery. When it comes to injection and hijacking, $\mu$Verum\cite{25_micro_intrusion_recovery} presents an interesting way of recovery, by reversing the actions done to the system.

\section{Discussion}
\label{section: discussion}

The reason for having such a disparity between detection and recovery methods can be explained by the fact that a lot of existing frameworks for microservice deployment already provide some fault tolerance features. Kubernetes provides node replication, and automatic CPU or memory scaling~\cite{12_automated_monitoring_lstm}. Service meshes like Istio provide authentication and authorization, service discovery, and traffic configuration~\cite{microservice_security_NIST}. If these mechanisms suffice for most business applications, leaving less incentive for researchers to explore. Additionally, recovery from some faults may involve changing business logic, which could be too difficult to automate.  

Very few of the surveyed papers contained well known datasets (such as AIOps Challenge 2020 or SockShop Dataset) and used platforms to inject specific faults into their testbeds. This makes it difficult to compare the quality of these methods against one another since there is no objective ground to compare them. Some papers use other past implementations to directly compare the performance, but since there is no established state of the art, the choices of which implementations are used differs between most papers. Future work could compare the discussed methods empirically, and propose a more objective framework to compare them.

Fault detection mechanisms can still be improved for effective fault recovery to take place. The majority of the surveyed papers were not able to accurately identify the root cause of a fault more than 90\% of the time on their own datasets. However, some faults seem to be easier to detect than others, so in the future, we propose that fault detection algorithms discuss in more detail their model's performance per each fault, rather than generalizing it for all injected faults.  

The most notable literature gap is runtime fault tolerance exploration of microservice specific components, such as service registry and monitoring mechanisms. This could explain why platforms that provide service registry already have built-in fault tolerance mechanisms, but no such mention was discovered in this survey. 

\subsection{Limitations}

The biggest limitation of this study is that the proposed threat model does not sufficiently cover the existing faults and vulnerabilities presented in microservices. This can be due to overgeneralization of faults, or ineffective modeling. In future work, a more layered approach for analysing faults could be employed, e.g., the model proposed by Yarygina and Bagge~\cite{microservice_security_layered_model}.
This survey did not fully follow a systematic approach for gathering primary papers for the survey. Because of this, important literature works may have been excluded during the first stages of the search. 

\section{Conclusion}
\label{section: conclusion}

This survey was set out to explore the current progress of runtime fault tolerance detection and recovery mechanisms. We found that those topics are still actively explored within the microservice architecture community, with the number of research papers on this sector increasing every year. Detection mechanisms have attracted more attention than recovery mechanisms, which we explain by the fact that most deployment frameworks already provide some form of fault recovery, or because fault recovery is too complicated to perform automatically. 

Using the faults that were identified during threat modeling, we showed that current literature covers most faults that relate to microservice performance, architecture and security. Currently, faults that relate to microservice specific infrastructure elements such as service discovery and monitoring mechanisms have not been sufficiently covered.

For detection mechanisms, there is no clear way of comparing their quality, since most of them create their own data sets for testing, or arbitrarily choose which past implementations to compare to. For future work, we suggest doing empirical analysis on these papers, and proposing a more objective framework, where these solutions can be compared to one another.

\printbibliography

\appendix 

\section{Search queries}
\label{section: queries}

Each digital library has its search query requirements and limitations. To address this, the general query was adjusted to fit these constraints. The general query can be seen in \autoref{table: general search query}. 

\begin{table}[b]
\caption{General search query}
\label{table: general search query}
microservice* AND (toleran* OR resilien* OR prevention OR forecasting OR removal OR recovery OR restoration OR availability OR dependability OR debugging OR performance prediction OR monitoring OR security) AND (byzantine OR crash OR fault OR fail* OR threat)
\end{table}

The subsequent queries are presented as they are interpreted by the search engine of the platform. These can be pasted within the advanced search of each website and should yield identical results to this survey. \\

\textbf{IEEE Xplore.}

\textit{("All Metadata":microservice*) AND ("All Metadata":toleran* OR "All Metadata":resilien* OR "All Metadata":prevention OR "All Metadata":forecasting OR "All Metadata":removal OR "All Metadata":recovery OR "All Metadata":restoration OR "All Metadata":availability OR "All Metadata":dependability OR "All Metadata":debugging OR "All Metadata":performance prediction OR "All Metadata":monitoring OR "All Metadata":security) AND ("All Metadata":byzantine OR "All Metadata":crash OR "All Metadata":fault OR "All Metadata":fail* OR "All Metadata":threat)}
\\

\textbf{ACM Digital Library.}

\textit{Abstract:(microservice* AND (toleran* OR resilien* OR prevention OR forecasting OR removal OR recovery OR restoration OR availability OR dependability OR debugging OR performance prediction OR monitoring OR security) AND (byzantine OR crash OR fault OR fail* OR threat)) OR Title:(microservice* AND (toleran* OR resilien* OR prevention OR forecasting OR removal OR recovery OR restoration OR availability OR dependability OR debugging OR performance prediction OR monitoring OR security) AND (byzantine OR crash OR fault OR fail* OR threat)) OR Keyword:(microservice* AND (toleran* OR resilien* OR prevention OR forecasting OR removal OR recovery OR restoration OR availability OR dependability OR debugging OR performance prediction OR monitoring OR security) AND (byzantine OR crash OR fault OR fail* OR threat))} \\

\textbf{Web of Science.}

\textit{(ALL=(microservice* AND (toleran* OR resilien* OR prevention OR forecasting OR removal OR recovery OR restoration OR availability OR dependability OR debugging OR performance prediction OR monitoring OR security) AND (byzantine OR crash OR fault OR fail* OR threat))) AND (PY==("2019" OR "2020" OR "2021" OR "2022" OR "2023" OR "2024") AND PUBL==("Elsevier" OR "Elsevier" OR "Springer Nature" OR "Mdpi" OR "Assoc Computing Machinery" OR "Wiley" OR "Taylor \& Francis" OR "Amer Geophysical Union" OR "NATURE PORTFOLIO" OR "Amer Chemical Soc" OR "Frontiers Media Sa" OR "Sage" OR "Copernicus Gesellschaft Mbh" OR "Amer Soc Mechanical Engineers" OR "Hindawi Publishing Group" OR "Oxford Univ Press" OR "Public Library Science" OR "SPRINGER INT PUBL AG" OR "Science Press" OR "Iop Publishing Ltd" OR "Scitepress" OR "Univ Chicago Press" OR "Wiley-Hindawi" OR "Amer Soc Civil Engineers" OR "Asce-Amer Soc Civil Engineers" OR "Asme" OR "Bmj Publishing Group" OR "E D P Sciences" OR "Igi Global" OR "Inderscience Enterprises Ltd" OR "Inst Engineering Technology-Iet" OR "Ios Press" OR "Keai Publishing Ltd" OR "Neural Information Processing Systems (Nips)" OR "Peerj Inc" OR "Seismological Soc Amer" OR "Soc Petroleum Eng" OR "Usenix Assoc" OR "AIP Publishing" OR "Acad Sci Czech Republic Inst Rock Structure \& Mechanics" OR "Amer Fisheries Soc" OR "Amer Inst Mathematical Sciences-Aims" OR "Amer Meteorological Soc" OR "Assoc Advancement Artificial Intelligence" OR "Budapest Tech" OR "Cambridge Univ Press" OR "Canadian Science Publishing" OR "China Univ Geosciences, Beijing" OR "Comsis Consortium" OR "FERDOWSI UNIV MASHHAD PRESS" OR "Fuji Technology Press Ltd" OR "Geological Soc Publishing House" OR "Graz Univ Technolgoy, Inst Information Systems Computer Media-Iicm" OR "Groupe Francias Geomorphologie" OR "Higher Education Press" OR "IEICE-INST ELECTRONICS INFORMATION COMMUNICATION ENGINEERS" OR "Ijcai-Int Joint Conf Artif Intell" OR "Indian Acad Sciences" OR "Iniestares, S.A." OR "Inst Navigation" OR "Int Journal Computer Science \& Network Security-Ijcsns" OR "Int Union Crystallography" OR "J A S S S" OR "JMLR-JOURNAL MACHINE LEARNING RESEARCH" OR "Korean Inst Communications Sciences (K I C S)" OR "Korean Soc Precision Eng" OR "Lippincott Williams \& Wilkins" OR "Ltd Georesursy" OR "Pleiades Publishing Inc" OR "Polish Maintenance Soc" OR "Polska Akad Nauk" OR "Princess Sumaya Univ \& SRSF" OR "Royal Soc Chemistry" OR "Science \& Information Sai Organization Ltd" OR "Sciendo" OR "Soc Brasileira Geologia" OR "Suranaree Univ Technology" OR "Systems Engineering \& Electronics, Editorial Dept" OR "Tech Science Press" OR "Thieme Medical Publishers" OR "Univ Federal Mato Grosso" OR "Univ Malaysia Pahang" OR "Univ Passo Fundo" OR "Univ Politecnica Madrid-Cepade" OR "V N Karazin Kharkiv Natl Univ" OR "Voronezh State Technical Univ" OR "Water Research Commission" OR "World Scientific") AND SJ==("COMPUTER SCIENCE" OR "ENGINEERING" OR "SCIENCE TECHNOLOGY OTHER TOPICS" OR "MATHEMATICAL COMPUTATIONAL BIOLOGY" OR "PEDIATRICS"))}

\end{document}